\documentclass[sigconf, screen, nonacm]{acmart}

\usepackage{fancyhdr}
\usepackage[normalem]{ulem}
\usepackage{amsmath}
\usepackage{mathtools}
\usepackage{tabularray}
\usepackage{xcolor,soul}
\usepackage{multirow}
\usepackage{subfig}
\usepackage{setspace}
\usepackage{cleveref}
\usepackage{enumitem}

\crefformat{section}{\S#2#1#3}
\crefformat{subsection}{\S#2#1#3}
\crefformat{subsubsection}{\S#2#1#3}

\newcommand\serverlessrefs{~\cite{implicationsfaasMICRO2019,catalyzerASPLOS2020,Ustiugov_2021,mementoMICRO2023,mahgoub2022wisefuse,shahrad2019architectural,sahraei2023xfaas,zhang2021faster,roy2022icebreaker,shahrad2020serverless,sebs,romero2021faa,mahgoub2021sonic,tzenetopoulos2023dvfaas,seqclock,sadeghian2023unfaasener,mahgoub2022orion,wen2022stepconf,eismann2021sizeless,moghimi2023parrotfish,joosen2023does,jonas2019cloud,stojkovic2023specfaas,bhasi2021kraken,lopez2020triggerflow,sampe2018serverless,mohan2019agile,stojkovic2024ecofaas,cai2024incendio,wu2024faasbatch,zhang2024jolteon,tzenetopoulos2025towards,xu2024faasmem,basu2024codecrunch,joosen2025serverless}}
\newcommand\microservicesrefs{~\cite{gan2018architectural,deathstarbench,zhang2024ursa,masouros2020rusty,kannan2019caliper,sriraman2018mutune,chen2019parties,patel2020clite,zhao2021understanding,nishtala2020twig,zhang2021sinan,hou2020ant}}
\newcommand\llminferencerefs{~\cite{vllmSOSP2023,touvron2023llamaopenefficientfoundation,murty2024bagelbootstrappingagentsguiding,jiang2023mistral7b,vaswani2023attention,openai2023gpt4,kakolyris2025throttll,tuli2024dynamo,fu2024serverlessllm}}
\newcommand\bioinformaticsrefs{~\cite{genomicsbench,albayraktaroglu2005biobench,kim2018grim,ghiasi2022genstore,cali2022segram,chaisson2012mapping,li2013aligning,li2009fast,ahmed2016comparison,alser2020accelerating,alser2017gatekeeper,appuswamy_ipdpsw2018,cali2020genasm,kim2017grim,graphaligner2020}}

\setcitestyle{numbers,sort&compress}
\usepackage{xspace}
\usepackage{color}
\usepackage{listings}
\usepackage{xcolor}

\definecolor{darkspringgreen}{rgb}{0.09, 0.45, 0.27}
\definecolor{denim}{rgb}{0.08, 0.38, 0.74}
\definecolor{darkolivegreen}{rgb}{0.33, 0.42, 0.18}
\definecolor{tangerine}{rgb}{0.95, 0.52, 0.0}
\definecolor{mahogany}{rgb}{0.75, 0.25, 0.0}

\definecolor{uglyyellow}{rgb}{0.99, 0.93, 0.0}
\definecolor{nbs}{rgb}{0.35, 0.31, 0.81}

\definecolor{darkred}{rgb}{0.75, 0.1, 0.1}

    \newcommand{\vnit}[1]{\textcolor{brown}{#1}}                

\newcommand{\revA}[1]{\textcolor{green}{#1}}

\definecolor{seagreen}{rgb}{0.18, 0.55, 0.34}

\definecolor{darkpink}{rgb}{0.88, 0.28, 0.54}
\definecolor{forestgreen}{rgb}{0.0, 0.27, 0.13}
\definecolor{amber}{rgb}{1.0, 0.49, 0.0}

\usepackage{tikz}

\newcommand{\squishlist}{
 \begin{list}{$\circ$}
  { \setlength{\itemsep}{0pt}
     \setlength{\parsep}{0pt}
     \setlength{\topsep}{3pt}
     \setlength{\partopsep}{0pt}
     \setlength{\leftmargin}{1em}
     \setlength{\labelwidth}{1em}
     \setlength{\labelsep}{0.5em} } }

\newcommand{\squishend}{
  \end{list}  }

\newcommand*\circled[1]{\tikz[baseline=(char.base)]{\node[shape=circle,fill,inner sep=0.5pt] (char) {\textcolor{white}{#1}};}}
\newcommand*\circledwhite[1]{%
  \tikz[baseline=(char.base)]{
    \node[
      shape=circle,
      draw,
      fill=white,
      inner sep=1pt
    ] (char) {\textcolor{black}{#1}};
  }%
}
\definecolor{denim}{rgb}{0.08, 0.38, 0.74}
\definecolor{azure(colorwheel)}{rgb}{0.0, 0.5, 1.0}
\definecolor{greenp}{rgb}{0.0, 0.65, 0.50}
\definecolor{peach}{rgb}{0.97, 0.51, 0.47}
\definecolor{darkmagenta}{rgb}{0.55, 0.0, 0.55}
\definecolor{royalblue(web)}{rgb}{0.25, 0.41, 0.88}
\definecolor{ao(english)}{rgb}{0.0, 0.5, 0.0}
\definecolor{ForestGreen}{RGB}{34,139,34}

\usepackage[colorinlistoftodos,prependcaption,textsize=scriptsize]{todonotes}

\usepackage[bottom]{footmisc}
\usepackage{dblfloatfix}
\usepackage{array}
 \usepackage{booktabs}

\definecolor{ufogreen}{rgb}{0.1, 0.6, 0.4}
\definecolor{ufogreen}{rgb}{0.1, 0.6, 0.4}

\usepackage{listings}
\usepackage{xcolor}
\usepackage[T1]{fontenc} 
\usepackage{zi4} 
\usepackage{caption} 
\usepackage{float} 

\lstdefinestyle{custompseudocode}{
  belowcaptionskip=1\baselineskip,
  breaklines=true,
  xleftmargin=\parindent,
  language=Python,
  showstringspaces=false,
  basicstyle=\small\ttfamily,
  keywordstyle=\bfseries\color{green!40!black},
  commentstyle=\itshape\color{purple},
  stringstyle=\color{orange},
  numbers=left,
  numberstyle=\scriptsize\color{black},
  numbersep=8pt,
  morekeywords={function}, 
  keywordstyle=[2]\bfseries, 
  escapeinside={*@}{@*}, 
  xleftmargin=2em, xrightmargin=0em, 
}

\newcommand{\kanellokcom}[1]{}

\newcommand{\vladcom}[1]{}

\definecolor{rcommon}{RGB}{0, 105, 180}
\definecolor{ra}{RGB}{34, 139, 34}
\definecolor{rb}{RGB}{196, 30, 58}
\definecolor{rc}{RGB}{25, 60, 170}
\definecolor{rd}{RGB}{220, 90, 10}

\providecommand{\revA}[1]{#1}
\providecommand{\revB}[1]{#1}

\providecommand{\revCQ}[1]{#1}
\renewcommand{\revA}[1]{{\color{black}{#1}}}
\renewcommand{\revB}[1]{{\color{black}{#1}}}

\renewcommand{\revCQ}[1]{{\color{black}{#1}}}

\newcommand{\bqnote}[1]{}
\newcommand{\aqnote}[1]{}
\newcommand{\cqnote}[1]{}
\newcommand{\dqnote}[1]{}
\newcommand{\commonqnote}[1]{}

\newcommand{\revahigh}[1]{\colorbox{ra!25}{#1}}
\newcommand{\revbhigh}[1]{\colorbox{rb!25}{#1}}
\newcommand{\revchigh}[1]{\colorbox{rc!20}{#1}}
\newcommand{\revdhigh}[1]{\colorbox{rd!25}{#1}}

\AtBeginDocument{%
  }

\setcopyright{none}

\begin{document}

\title[Valinor: Fast, Energy-Efficient and Programmable Physical Memory Allocation]{Valinor: Architectural Support for Fast, Energy-Efficient \\and Programmable Physical Memory Allocation}

\author{%
  Konstantinos Kanellopoulos\textsuperscript{1} \quad
  Spiros Galanopoulos\textsuperscript{1} \quad
  Konstantinos Sgouras\textsuperscript{1} \quad
  Vlad-Petru Nitu\textsuperscript{1} \quad \\
  Ilias Papalamprou\textsuperscript{2} \quad
  Andreas Kosmas Kakolyris\textsuperscript{1} \quad
  Rahul Bera\textsuperscript{1} \quad
  Dimosthenis Masouros\textsuperscript{2} \quad \\
  Dimitrios Soudris\textsuperscript{2} \quad
  Onur Mutlu\textsuperscript{1}%
  \vspace{2mm}
}
\affiliation{%
  \institution{\textsuperscript{1}ETH Zürich \quad
    \textsuperscript{2}National Technical University of Athens}
  \country{}
}

\begin{abstract}
Physical memory allocation is a core OS mechanism that establishes on-demand virtual-to-physical mappings. In today’s systems, this mechanism remains software-centric: each page fault traps into the kernel, triggering pipeline flushes, stalls, and a long sequence of allocation steps that can cost tens of thousands of cycles. While such overheads were tolerable for long-running, monolithic applications, they impose substantial latency and energy penalties on emerging short-lived workloads with highly ephemeral allocation patterns (e.g., serverless, microservices). For these workloads, even minor faults, those not requiring disk I/O, now account for up to 54\% of runtime and up to 40\% of system energy. Prior work has attempted to offload allocation to hardware to avoid traps and context switches, but existing designs suffer from two key limitations: (i) they sacrifice important data-placement optimizations (e.g., bank-level page coloring) in the name of speed, and (ii) they rely on fixed-function logic that cannot adapt to new policies or evolving hardware conditions.

We present \textbf{Valinor}, a hardware–OS cooperative memory allocation substrate that combines software flexibility with hardware-class performance. Valinor introduces a programmable hardware allocation engine capable of executing OS-supplied allocation libraries, i.e., compact routines that implement allocation policies at close to fixed-hardware speed. This design enables low-latency allocation while accommodating diverse policies, including short-lived object allocators, integrity-enforcing mechanisms, and hardware-telemetry-guided placement strategies.
We implement Valinor on a BOOM RISC-V soft-core running Linux and in a full-system simulator. On a real hardware prototype, Valinor accelerates allocation by 17× and improves end-to-end performance by 16\%, while reducing energy consumption by up to 8\%. Using full-system simulation, we further explore the design space of the programmable allocation engine (PAE) and evaluate six distinct allocation libraries, demonstrating that Valinor can efficiently support diverse policies while achieving hardware-class performance without sacrificing programmability.
\end{abstract}

\maketitle

\pagestyle{plain}


\section{Introduction}
\label{sec:introduction}
Physical memory allocation is a fundamental OS service that has a growing impact on the performance and energy efficiency 
in modern computing systems~\cite{old_vm1,old_vm2,old_vm3,old_vm4}. Each allocation typically involves handling a page fault in software: the processor traps into the operating system, which first assigns a physical frame to establish the virtual-to-physical mapping.

As we demonstrate in \S\ref{sec:motivation} and as recent studies show~\cite{mementoMICRO2023,hbdpISCA2020,minorfaultTACO2022,Ustiugov_2021}, allocation-related OS activity in modern short-running user-interactive workloads can account for up to 54\% of their execution time and for up to 18\% of their energy footprint. 

To address the limitations of existing purely software- and hardware-based allocation schemes~\cite{minorfaultTACO2022,hbdpISCA2020,mementoMICRO2023}, we propose \emph{Valinor}, a hardware–OS cooperative memory allocation substrate that combines the flexibility of software-based allocators with the low-latency and energy-efficiency of hardware-based allocation. Valinor introduces a programmable allocation engine (PAE) that runs compact software-defined allocation libraries directly in hardware. The OS loads and configures these libraries using process-level directives, allowing page faults to be resolved without entering the kernel’s slow fault-handling path. Unlike prior hardware allocators that implement a fixed, greedy strategy, Valinor enables the OS to select from a diverse collection of allocation libraries, each tailored to a different optimization goal or hardware environment. This way, Valinor enables various intelligent memory allocation policies while eliminating the costly overheads of the software page fault path, context switching, and pipeline flushes during allocation.

\textbf{Mechanism Overview.} Applications provide coarse-grained (per-process) or fine-grained (per-memory object) directives that allow the OS to associate specific memory regions with an allocation library. The OS then configures the hardware allocation engine accordingly, enabling hardware-level execution of that library’s policy. When a core encounters a page fault, the MMU determines whether the faulting memory object is bound to a library and whether it is a candidate for hardware-based handling. If so, the request is forwarded to the PAE, which invokes the library’s routines. The library itself decides whether it can (or should) allocate the object: on success, it selects a physical page, updates its metadata, and returns the resulting PTE directly to the MMU, completing the allocation \emph{entirely in hardware}. If the library declines to allocate, Valinor transparently falls back to the standard OS page-fault handler.

At a high level, Valinor provides three key benefits:

\noindent\textbf{(1) Hardware-class latency and energy efficiency.}
In latency-sensitive environments (e.g., serverless functions), Valinor bypasses the traditional software page fault path and satisfies many allocation requests directly via the hardware allocation engine. This way, Valinor avoids the overheads of context switches and pipeline flushes and executes allocator logic on a lightweight, energy-efficient in-order pipeline, substantially reducing both latency and energy consumption during allocation (see \S\ref{sec:results}).

\noindent\textbf{(2) Rich programmability and library extensibility.}
Valinor is not a fixed-function allocator. Instead, it exposes a programmable interface that allows the OS to load specialized allocation libraries into hardware. System developers can supply a wide library collection, including fast allocation, energy-efficient allocation,
and more (see \S\ref{sec:design:specialized_libraries}), without modifying the hardware. Libraries can manage their own metadata and define placement heuristics. This flexibility preserves the expressive power of software-based allocators while achieving high performance (see \S\ref{sec:results}).

\noindent\textbf{(3) Adaptivity to microarchitectural state.}
The PAE resides in the memory controller and maintains direct visibility into low-level signals (e.g., DRAM bank conflicts). This way, allocation libraries incorporate fine-grained hardware feedback to make adaptive placement decisions. This enables optimizations, such as steering allocations away from congested banks, that are infeasible in conventional software-based allocators (see \S\ref{sec:results}).

\textbf{Key Results.} We prototype Valinor in a BOOM RISC-V soft-core running Linux (i.e., on a Xilinx ZCU106 FPGA~\cite{zcu106}) to evaluate its feasibility and performance. We also implement Valinor in a full-system simulator (Virtuoso~\cite{virtuoso_arxiv}+Sniper~\cite{sniper}) to perform a wide design space exploration for the allocation engine's architecture (e.g., in-order core vs OoO vs CGRA) and to evaluate multiple allocation libraries.
\aqnote{Q9}\revA{We evaluate six different memory allocation libraries: (i)~a \emph{low-latency} allocator that organizes memory in a hash-based manner; (ii)~a \emph{per-page-MAC} and (iii)~a \emph{Merkle-tree} integrity-enforcing library, both securing virtual-to-physical mappings; (iv)~a \emph{speculation-based} library that accelerates allocation; (v)~a \emph{telemetry-driven} library that adapts data placement to runtime microarchitectural conditions to reduce interference; and (vi)~a \emph{tier-aware} library that places pages across DRAM, NUMA, and CXL tiers.}

Our evaluation yields four key results. First, on the RISC-V prototype, Valinor reduces minor page-fault handling time by \textbf{17$\times$} compared to Linux, achieving \textbf{16\%} average end-to-end speedups and \textbf{5\%} average energy savings. Despite using a simple in-order programmable core, Valinor remains within \textbf{3\%} of fixed-function hardware while preserving full flexibility. Second, we evaluate Valinor across six allocation libraries and demonstrate that it can support low-latency, integrity-enforcing, speculative, and adaptive placement policies without any hardware redesign. Notably, the speculation-based library hides \textbf{over 99\%} of page-fault latency. Third, Valinor’s telemetry-driven library uses real-time microarchitectural signals to steer allocations away from congested DRAM banks, restoring co-located latency-sensitive performance to within \textbf{1--2\%} of isolated execution while still accelerating allocation-sensitive workloads by \textbf{1.25$\times$}, outperforming both software-only and greed-for-speed hardware allocators. Finally, Valinor introduces modest hardware cost: in an 8-core BOOM CPU design, the PAE adds only \textbf{$<$1.5\%} area and \textbf{1.5--1.8\%} static power. These costs are small relative to the substantial reductions in page-fault latency, energy, and interference that Valinor provides.
In this work, we make the following contributions:

\begin{itemize}[leftmargin=*]

\item We perform an extensive characterization of the Linux page-fault path, quantifying its latency, energy cost, and bottlenecks across emerging workloads. This analysis motivates the need for hardware-assisted allocation.

\item We propose \textbf{Valinor}, a hardware-OS cooperative memory allocation scheme that combines the flexibility of software-based with the  efficiency of hardware-based allocation. Valinor introduces a PAE that allows the CPU to offload allocation requests directly to hardware, eliminating costly kernel traps and context switches while supporting diverse allocation policies with different optimization goals.

\item We demonstrate a broad set of allocation policies enabled by Valinor’s programmable hardware–software interface, including low-latency allocation for short-lived objects, adaptive placement guided by microarchitectural telemetry (e.g., DRAM bank pressure), and integrity-checking of virtual-to-physical mappings.

\item We implement Valinor in both a Linux prototype on a BOOM RISC-V soft-core and a full-system simulator (Virtuoso+Sniper). Across short-lived and placement-sensitive workloads, Valinor achieves 17$\times$ reduction in page-fault latency, delivers  \textbf{16\% end-to-end speedups on average}, and provides up to \textbf{8\% energy savings}, while adding less than \textbf{1.5\%} area to a 8-core BOOM processor. We will open-source our infrastructure to enable further use in the community.

\end{itemize}

\section{Background}
\label{sec:background}

Modern processors implement \textbf{address translation} to bridge the gap between the virtual address space exposed to software and the physical address space used by hardware. Each memory reference issued by the CPU must be translated from a Virtual Address (VA) to a Physical Address (PA) before it can access memory. Address translation is performed by looking up per-process page tables managed by the operating system.
Page tables define the mapping between virtual pages and physical frames at a fixed granularity (typically 4\,KB). Each entry, called a Page Table Entry (PTE), stores the physical frame number and a set of metadata bits such as the \textit{Present} bit, access permissions (read, write, execute), and state bits (accessed and dirty). A cleared Present bit indicates that the virtual page is not currently mapped in physical memory. In such cases, any access to that address results in a \emph{page fault} exception, causing the processor to trap into the kernel’s page fault handler for resolution.

To minimize translation overhead, processors maintain a hardware unit called the Memory Management Unit (MMU) which is responsible for address translation. The MMU employs small hardware caches known as the Translation Lookaside Buffers (TLB) to store recently-accessed PTEs.
On every memory access, the MMU first queries the TLB. A \textbf{TLB hit} allows the translation to complete with low latency. A \textbf{TLB miss}, however, forces the hardware to perform a \textbf{page table walk}, during which the MMU traverses the multi-level page table to locate the corresponding PTE. If the PTE is found and marked as present, the MMU updates the TLB with the new translation and resumes execution. If the PTE is not present, a page fault is raised, transferring control to the OS' page fault handler routine.

\subsection{Demand Paging and Page Fault Handling}
The Memory Management Unit (MMU) handles address translation and uses Translation Lookaside Buffers (TLBs) to cache recent PTEs. On each memory access, the MMU checks the TLB. A hit resolves quickly. A miss triggers a page table walk (PTW) through the multi-level page table. If the PTE is found and present, the TLB is updated and execution resumes. Otherwise, a page fault invokes the OS.

\begin{figure}[h]
    \centering
    \includegraphics[width=\linewidth]{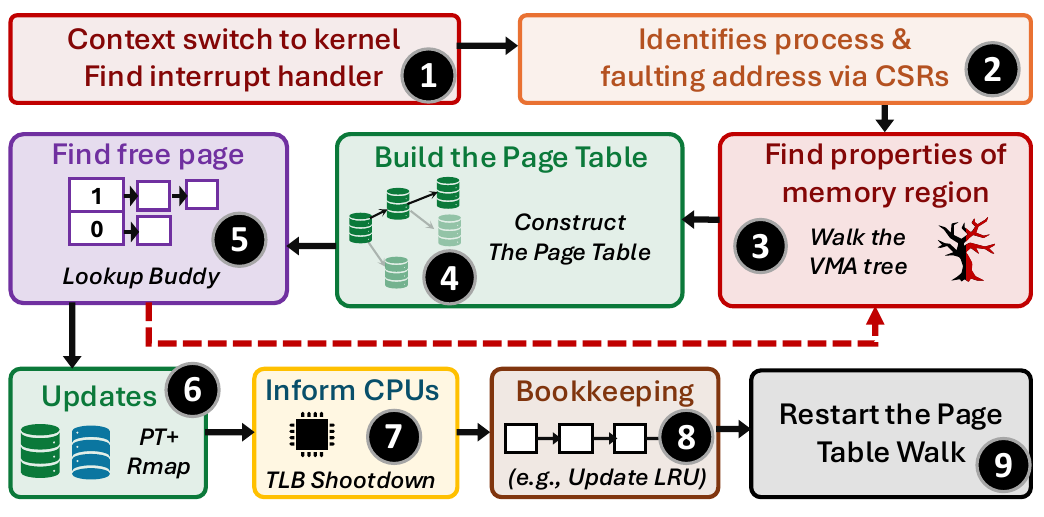}
    \caption{Resolving an anonymous, minor page fault in Linux}
    \label{fig:page_fault_handling}
\end{figure}

Modern operating systems (e.g., Linux) use \textbf{demand paging} to manage memory: pages are brought into physical memory only when first accessed, reducing startup latency and allowing workloads to exceed available DRAM.
When a process accesses a virtual address (VA) that is not backed by physical memory, the hardware triggers a \textbf{page fault}, and the kernel resolves it by allocating memory, updating metadata, and resuming execution.
Fig.~\ref{fig:page_fault_handling} shows the high-level steps for the simplest case: a \textbf{minor fault on an anonymous page}. When the fault occurs, control transfers to the kernel (\circled{1}), which identifies the faulting process and address (\circled{2}). The kernel locates the faulting region by walking the \textbf{Virtual Memory Area (VMA)} tree (\circled{3}). VMA entries describe contiguous regions of a process’s virtual address space (VAS) and their attributes.
For a newly touched anonymous region, the kernel allocates a physical frame by querying the buddy allocator~\cite{buddy} (\circled{5}) and then creates or updates the \textbf{page table entries (PTEs)} to establish the mapping (\circled{4}, \circled{6}). It updates the VMA information as needed and propagates the new mapping to CPUs, potentially issuing TLB shootdowns. Finally, the kernel performs bookkeeping such as updating the LRU list (\circled{8}) before returning to user space, where the PTW is restarted~(\circled{9}).
Modern operating systems (e.g., Linux) employ \textbf{Demand Paging} to manage memory efficiently. Rather than loading all pages of a process into memory at startup, pages are brought into physical memory only when they are first accessed. This design reduces startup latency and allows the system to execute workloads that exceed the available physical memory. When a process references a virtual address that does not currently have a valid page table entry, the hardware triggers a \textbf{page fault}. The kernel’s page fault handler is then responsible for resolving the fault by allocating memory, updating the relevant data structures, and resuming execution.

Fig.~\ref{fig:page_fault_handling} illustrates the high-level steps involved in handling a page fault in Linux, focusing on the simplest case: a \textbf{minor fault on an anonymous page}. Such a fault occurs when the accessed page does not yet exist in memory but can be immediately satisfied by allocating a new physical page, without requiring disk I/O.
When the fault occurs, control is transferred from user space to the kernel (\circled{1}), where the interrupt handler identifies the faulting process and the virtual address that caused the exception (\circled{2}). The kernel then determines the properties of the faulting memory region by walking the \textbf{Virtual Memory Area (VMA)} tree (\circled{3}). Each process maintains a set of VMAs that describe contiguous regions of its virtual address space with uniform attributes such as access permissions and backing source (e.g., anonymous or file-backed). The faulting address is located within this structure to identify the appropriate handling strategy.

In the case of a newly accessed anonymous region, the kernel proceeds to allocate a new physical frame. It first locates a free page by querying the buddy allocator~\cite{buddy} (\circled{5}) and retrieving a pre-zeroed page,
the kernel then constructs or updates the corresponding \textbf{page table entries (PTEs)} to establish the virtual-to-physical mapping (\circled{4}, \circled{6}). At the same time, the VMA tree is updated to reflect the new mapping. The updated mapping is propagated to the relevant CPUs, performing TLB shootdowns if necessary (\circled{7}), ensuring memory consistency across cores. Finally, the kernel performs internal bookkeeping, such as updating the least-recently-used (LRU) list for page replacement (\circled{8}), before returning control to user space.

This sequence represents the fundamental fast path of page fault resolution. More complex cases, such as copy-on-write (CoW) or faults requiring disk access, extend this workflow with additional stages but rely on the same underlying mechanisms. Table~\ref{tab:fault_types} summarizes the most common fault categories.

\begin{table}[h]
\centering
\footnotesize
\caption{Classification of Page Faults in Linux}
\label{tab:fault_types}
\renewcommand{\arraystretch}{1.15}
\setlength{\tabcolsep}{3pt}
\begin{tabular}{|m{0.8cm}|m{3.0cm}|m{4.1cm}|}
\hline
\textbf{Type} & \textbf{Example Scenario} & \textbf{Handling Description} \\
\hline
\multirow{3}{=}{\centering \textbf{Minor (no disk I/O)}} &
Lazy allocation of anonymous memory &
Allocate a zero-filled physical page and update the page table. \\
\cline{2-3}
& Copy-on-Write (CoW) &
Allocate a new page, copy the data, and update the PTE with write permissions. \\
\cline{2-3}
& File-backed page already in page cache &
Map the existing cached page into the process’s address space. \\
\hline
\multirow{2}{=}{\centering \textbf{Major}} &
File-backed page not in the page cache &
Fetch the file block from disk into the page cache, then update the page table. \\
\cline{2-3}
& Anonymous page swapped out to disk &
Read the page from swap space and restore the mapping. \\
\hline
\end{tabular}
\end{table}

\revA{
\subsection{Physical page allocation.}
To obtain a free physical frame, the kernel queries the \emph{buddy allocator}, which keeps each memory zone's free pages on per-order free lists of power-of-two contiguous blocks (an order-$n$ block spans $2^n$ pages), protected by a per-zone spinlock \texttt{zone->lock}.
To keep this lock off the fast path, 4KB pages are cached in per-CPU lists (\emph{pcplists}).
Once these are drained, allocations fall back to the buddy free lists and contend on \texttt{zone->lock}.
Under fragmentation, no contiguous high-order block may be available, so the allocator runs \emph{compaction}, migrating movable pages to coalesce free blocks.
This significantly increases the fault's critical path, due to contention on \texttt{zone->lock} and the LRU lists.
}
\aqnote{Q7}

\vspace{-3mm}
\section{Motivation}
\label{sec:motivation}

\subsection{The Shifting Nature of Modern Workloads}

Traditional applications such as scientific simulations~\cite{Plimpton2006,Tramm2014,rcnn-weather,dongarra_hpcg2015},
database servers~\cite{mongodb,redis,memcached,tpcc,tpch}, large-scale graph analytics~\cite{ligra,ham2016graphicionado,pagerank1999,gmsMaciej,graph500}
and genomics workloads~\bioinformaticsrefs\ are typically long-running and operate over large memory footprints.
These workloads hide the overheads of software-based memory allocation by amortizing them over the billions of instructions
they execute.

In contrast, the modern computing landscape is increasingly dominated by user-interactive, \emph{short-lived workloads}.
Examples include \emph{serverless functions executed in FaaS platforms}~\serverlessrefs,
\emph{microservices}~\microservicesrefs, and
\emph{large language model (LLM) inference queries}~\llminferencerefs,
perform limited computation per request, exhibit bursty request patterns, are stateless and often terminate within a hundred milliseconds.
These workloads allocate and release memory frequently and unpredictably, and they may touch only a small number of pages per request.
Because each request's lifetime is short, there is little opportunity to amortize the cost of page allocation and initialization.
As a result, even \emph{minor} page faults and associated physical page allocation routines can become a first-order
component of tail latency and energy consumption.

\subsection{Examples: Serverless and Microservices}
In FaaS platforms, a function invocation includes initialization, execution, and keep-alive~\cite{xu2024faasmem,tzenetopoulos2025towards}. If a function cannot stay warm, snapshotting reduces cold starts but restores pages with CoW permissions, so first writes trigger CoW minor faults. Processing a new request that executes a function incurs additional faults by triggering first-touch accesses and allocating dynamic memory.
For example, a JSON-processing pipeline repeatedly touches new intermediate pages with little data reuse, generating many minor faults~\cite{langdale2019parsing}. Microservices show similar behavior. At startup they map large anonymous regions. Per-request handlers then deserialize inputs, build metadata, and emit responses. Each step touches fresh pages\microservicesrefs. Although these services are long-lived, the short per-request work cannot amortize the resulting faults.

\subsection{Cost of Software-Based Memory Allocation}

\begin{figure}[h!]
  \vspace{-2mm}
  \centering
  \includegraphics[width=1.0\linewidth]{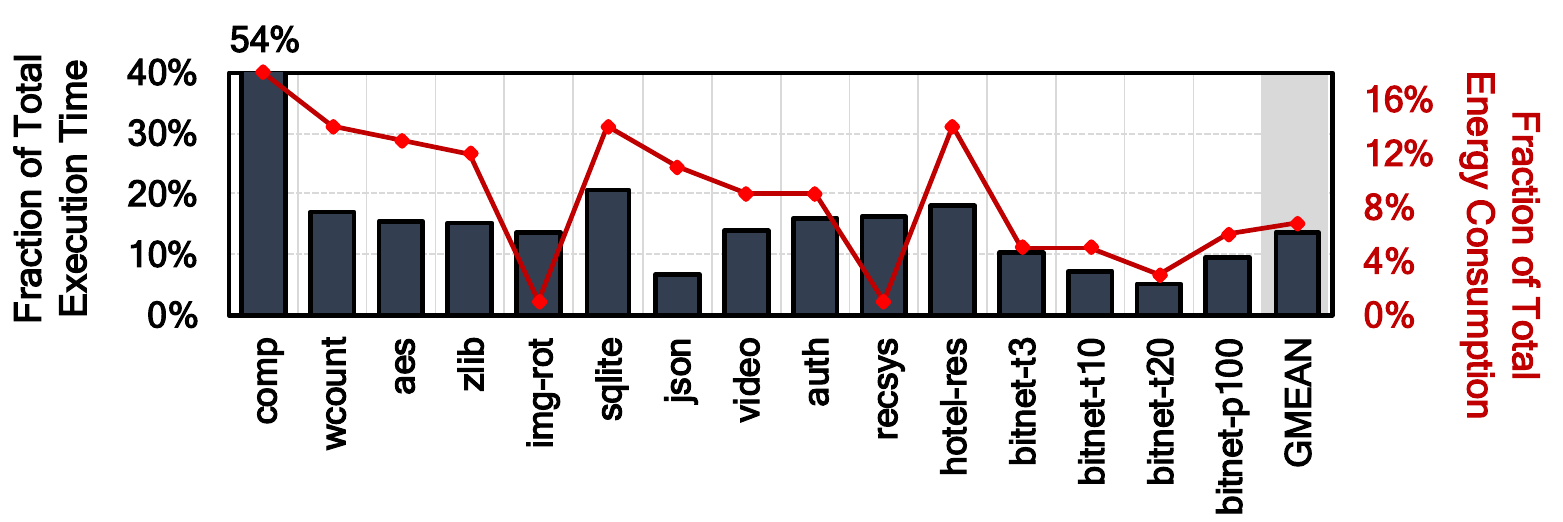}
  \vspace{-2mm}
  \caption{Fraction of total execution time and energy spent in minor page fault handling across diverse workloads, measured
  on a real server-grade CPU.}
  \label{fig:minor-latency-energy}
  \vspace{-2mm}
\end{figure}

To quantify the impact of software-based memory allocation on modern workloads,
we perform detailed profiling on a real Intel Xeon server using  DeathStarBench\cite{deathstarbench}, vSwarm\cite{vswarm}, and Bitnet~\cite{wang20241bitaiinfra11} (see \S\ref{sec:methodology}).
Fig.~\ref{fig:minor-latency-energy} reports the fraction of total execution time and energy (measured with Intel RAPL~\cite{david2010rapl}) spent in minor page-fault routines across our workloads. We make two observations.
First, minor faults consume on average 13\% (up to 54\%) of execution time and 7\% (up to 18\%) of total energy.
Second, even compute-intensive applications such as LLM inference in Bitnet
still spend \textbf{up to 11\% of total time} in minor faults.

\subsection{Dissecting Minor Page Fault Overheads}

To understand the source of these overheads, we dive deeper into the
performance characteristics of minor page faults.

\begin{figure}[h!]
  \centering
  \includegraphics[width=1.0\linewidth]{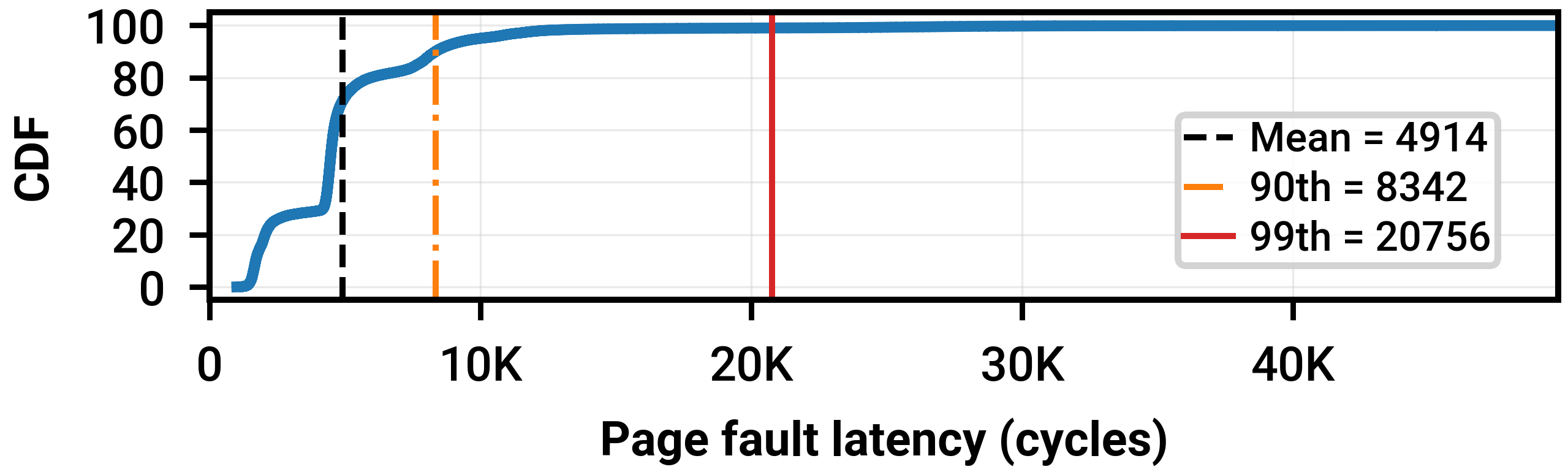}
  \vspace{-5mm}
  \caption{CDF of execution latency of minor page faults.}
  \label{fig:minor-latency-cdf}
    \vspace{-2mm}
\end{figure}

\textbf{Latency distribution.}
We develop an eBPF-based tracing tool that records the latency and retired instructions of minor page-fault paths.
First, our tool reveals that in these workloads, over 90\% of the total minor page faults occur due to anonymous pages.
Fig.~\ref{fig:minor-latency-cdf} shows
the CDF of fault latency in CPU cycles. The mean is ~4.9K cycles. The tail is long. The 90th percentile is above 8.3K cycles, and the 99th percentile is over 20K cycles.

These outliers arise from extra cache misses in the allocator or page tables, or contention on shared structures (e.g., zone locks, LRU lists).
Such latencies significantly degrade performance and inflate tail latency for short-lived functions.

\begin{figure}[h!]
  \vspace{-2mm}
  \centering
  \includegraphics[width=1.0\linewidth]{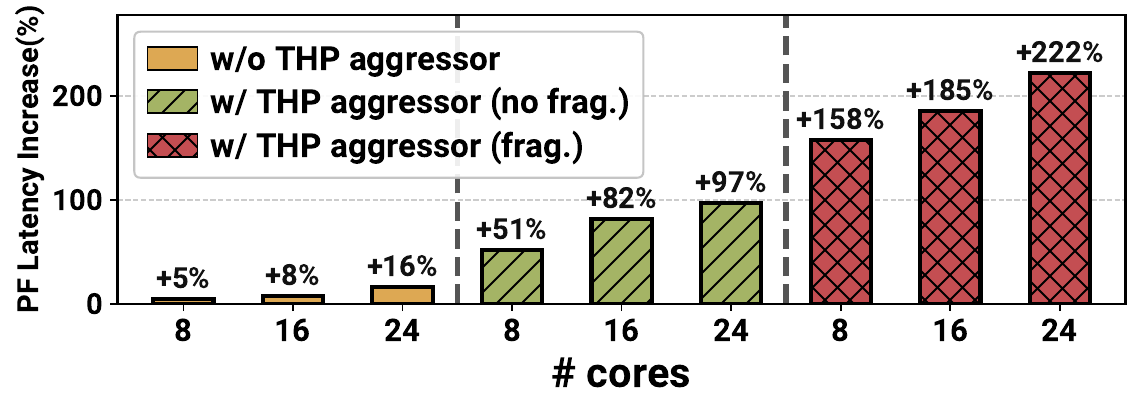}
  \vspace{-2mm}
  \caption{Page fault latency overhead with and without THP aggressors, both with and without memory fragmentation}
  \label{fig:thp-aggressor-pf-latency-increase}
  \vspace{-2mm}
\end{figure}

\textbf{Co-located workloads.}
To quantify contention on kernel-shared structures, we augment our eBPF-based tracing tool to measure how a victim workload's page fault latency varies depending on co-located aggressors.
Fig.~\ref{fig:thp-aggressor-pf-latency-increase} reports the page fault latency increase normalized to the victim running in isolation.
The victim is pinned on CPU~0. Each aggressor occupies a separate core (e.g., 8~cores = 1~victim + 7~aggressors).
We use three aggressor categories. \textbf{(1)~No THP aggressor}: allocates 512\,MB and touches every 4\,KB page once in random order, triggering one fault per page. \textbf{(2)~THP aggressor (no fragmentation)}: allocates 256\,MB and touches memory at a 2\,MB stride to obtain huge pages. \textbf{(3)~THP aggressor (fragmentation)}: allocates 1\,GB, touches every 4\,KB page, then releases physical memory via \texttt{MADV\_DONTNEED} to create a checkerboard-like fragmentation pattern. We then run aggressor~(2) to mimic allocation in a highly fragmented system, which triggers memory compaction during fault handling.

We make three key observations.
First, no-THP aggressors inflate the victim's page fault latency minimally (5\%, 7\%, and 17\% for 8, 16, and 24 cores), as they hold shared locks only briefly.
Second, THP aggressors without fragmentation increase latency by 51\% to 97\% (7 to 23 aggressors), because they hold \texttt{zone->lock} for many cycles, serializing the victim's fault handler.
Third, under memory fragmentation, ubiquitous in cloud-native systems~\cite{hotmem2024}, the increase reaches 163\% to 258\%, as compaction on the buddy allocator's critical path extends \texttt{zone->lock} hold times further.

\begin{figure}[h!]
  \vspace{-2mm}
  \centering
  \includegraphics[width=1.0\linewidth]{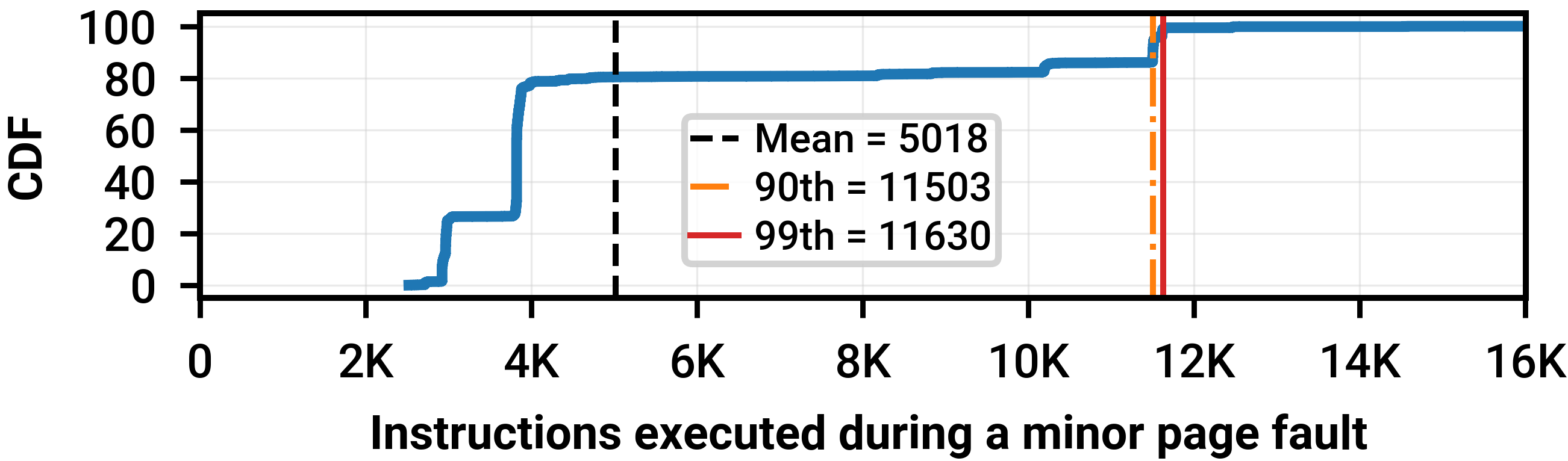}
    \vspace{-5mm}
  \caption{CDF of number of instructions executed during minor page faults.}
    \vspace{-2mm}
  \label{fig:minor-inst-detailed}
\end{figure}

\textbf{Instruction count.}
Fig.~\ref{fig:minor-inst-detailed} presents the CDF of the number of retired instructions per minor page fault, obtained from hardware performance counters.
Each fault retires more than 5K instructions on average, with the tail reaching over 11K instructions.
Such a high instruction count stresses the out-of-order core and contributes to both high latency and high energy consumption.

\begin{figure}[h!]
  \centering
  \includegraphics[width=1.0\linewidth]{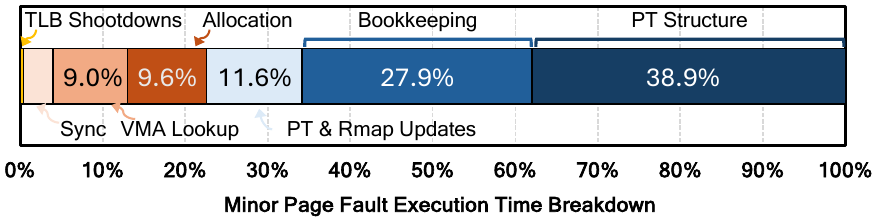}
  \caption{Minor page fault latency breakdown.}
  \label{fig:minor-fault-breakdown}
\end{figure}

\textbf{Software pipeline composition.}
Fig.~\ref{fig:minor-fault-breakdown} illustrates the breakdown of minor page-fault latency across its key software components:
(i) page-table operations,
(ii) bookkeeping (e.g., LRU and cgroup accounting),
(iii) page-table and rmap updates,
(iv) physical page allocation,
(v) VMA lookup and permission checks
(vi) synchronization and interrupt handling, and
(vii) TLB shootdowns.
We measure the execution time of over 150 kernel functions involved in minor faults using an \texttt{ftrace}-based~\cite{ftrace} function-level tracer and aggregate them into these categories.
Fig.~\ref{fig:minor-fault-breakdown} reports the average breakdown: constructing page-table entries and bookkeeping dominate (38\% and 27\%), and no single component determines total cost. Thus, minor-fault latency arises from the combined cost of many tightly coupled steps.

\section{Limitations of Prior Work}
Prior works proposed accelerating page allocation by offloading parts of memory management to hardware, avoiding software fault-handling overheads~\cite{minorfaultTACO2022,hbdpISCA2020,mementoMICRO2023}. Tirumalasetty et al.~\cite{minorfaultTACO2022} extend the MMU with a hardware-managed free-page queue that the kernel replenishes in the background, allowing minor anonymous faults to be satisfied without kernel traps. Lee et al.~\cite{hbdpISCA2020} show that faster storage increases the relative cost of software fault handling and introduce a HW engine that services major faults entirely in hardware. By removing context switches, pipeline flushes, and software overheads, these works show that hardware-based fault handling can reduce allocation overheads.

\aqnote{Q8}\revA{However, existing hardware schemes face two \emph{orthogonal} limitations: (i) \textit{greed-for-speed allocation} concerns \emph{which} page is chosen, while (ii) \textit{hardwired policies} concerns \emph{whether} the policy can change.
Under (i), simple heuristics (e.g., picking the first free page) optimize only for latency and are policy-blind: OS-enforced placement such as page coloring~\cite{pagecoloring}, tiered memory allocation~\cite{tiered,tldram}, and Rowhammer-aware placement~\cite{citadel} cannot be honored when hardware blindly takes the next free page, increasing contention or violating partitioning (e.g., bank-level isolation) and offsetting the latency gains (see \S\ref{sec:results:sim_interference}).
Under (ii), the logic is frozen into fixed-function units that cannot adapt to changing objectives, workloads, or memory topologies.
These are two distinct dimensions: a fixed-function unit with a smart policy still suffers (ii), while a programmable unit running a naive policy still suffers (i), so fixing one does not fix the other.
Valinor resolves both, enabling fast, policy-aware decisions without sacrificing programmability via allocation libraries that run on the PAE.}

\section{Valinor: Design Overview}

\begin{figure}[h!]
    \centering
    \includegraphics[width=1.0\linewidth]{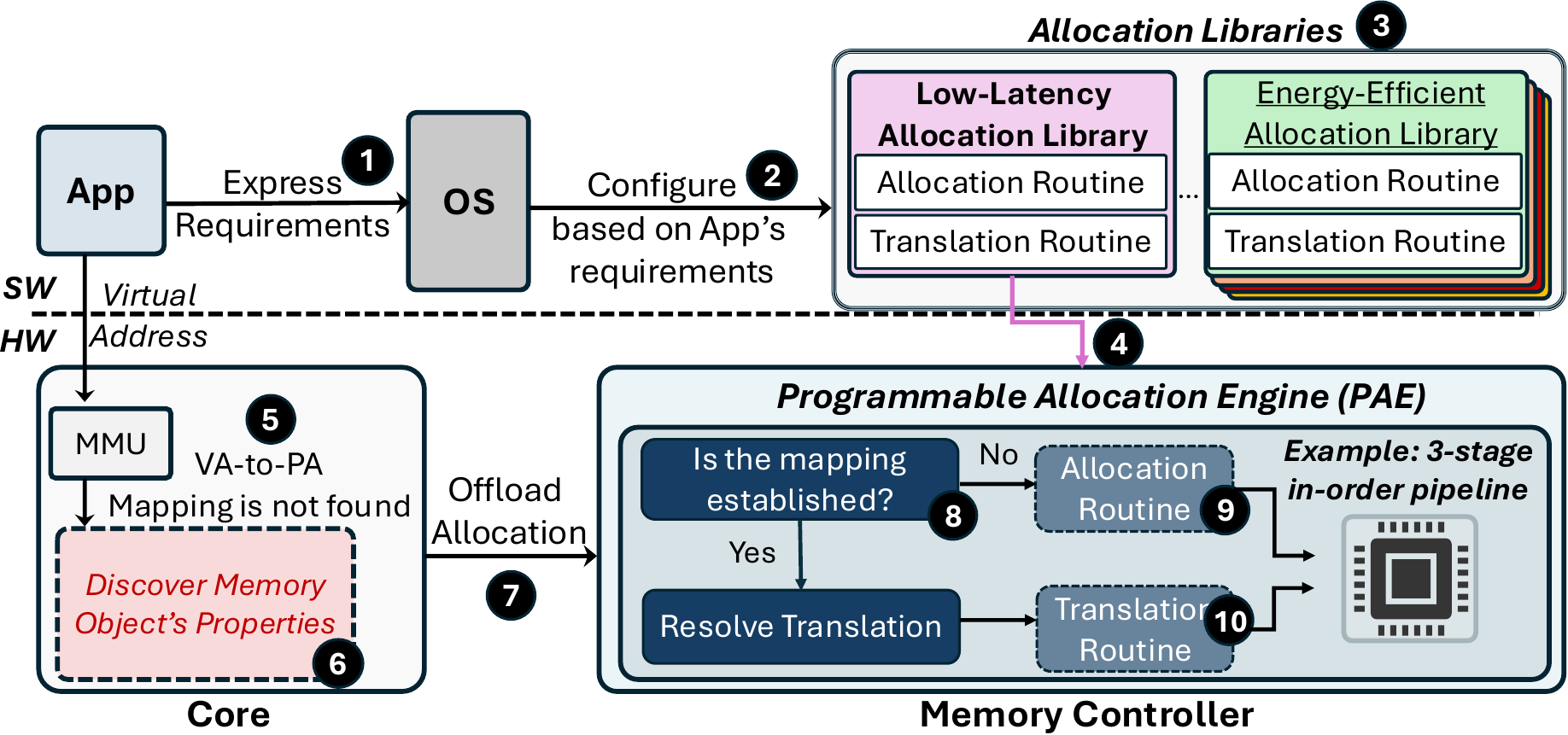}
    \caption{Valinor's architecture and integration with the OS.} 
    \label{fig:valinor_overview}
\end{figure}

We propose \textbf{Valinor}, a new hardware-OS cooperative memory allocation scheme that maintains the flexibility of software-based allocators while providing the low latency and energy efficiency of hardware-based ones.
Valinor’s architecture centers around a programmable hardware allocation engine (PAE) that executes allocation libraries inside hardware.
Fig.~\ref{fig:valinor_overview}
shows an overview of Valinor's end-to-end flow, which consists of
\textbf{four cooperating components}, each bringing its own key benefit.

\textbf{Application/OS Interface \textit{(Programmability)}.}
Applications explicitly express their allocation requirements, either coarsely for the entire process or finely for specific memory objects, by issuing special directives~\circled{1}. These directives are conveyed to the OS via system calls, allowing applications to specify which library should govern which memory objects. The OS interprets these directives and selects the appropriate allocation library~\circled{2}.  This interface provides programmability: different regions of an application can be bound to different policies without modifying hardware, and the OS retains full authority over which library is used where.

\textbf{Allocation Libraries \textit{(Custom \& Intelligent Policies)}.}
Each allocation library defines a complete memory-management policy through a
small set of functions. Libraries encapsulate policy (e.g., low-latency) and maintain their own metadata
(e.g., hash-based sets, private free lists). This enables policy specialization that existing hardware allocators can not express.  
To activate a library, the OS loads its code and metadata into memory~\circled{3} and configures the PAE by passing the required pointers~\circled{4}.

\textbf{MMU Extensions for Library Binding \textit{(Eligibility)}.}
When the MMU encounters a TLB miss \circled{5}, the page-table walker (PTW)
determines whether the faulting address belongs to a library-bound region.  
Valinor augments the PTW with a lightweight filtering structure that
extracts the properties of the memory object \circled{6} (e.g., anonymous versus file-backed). 
This filter determines in hardware whether there is a library bound to the object (i.e., hardware allocation is applicable~\circled{7}).  
This preserves low-latency allocation by preventing unnecessary traps into the OS, while ensuring libraries only see faults they are expected to handle.

\textbf{Programmable Allocation Engine (PAE) \textit{(Hardware-Class Latency + Adaptive Placement)}.}
If an object is associated with a library, the PTW forwards the request to the PAE~\circled{7}, which executes the library’s \texttt{translate} routine to check whether the page is already mapped~\circled{8}/\circled{10}.
If not, it runs the library’s \texttt{alloc} routine~\circled{9}, which may choose a physical page using policy metadata and hardware-visible state (e.g., DRAM bank load). On success, the PAE returns the physical page number (PPN) to the PTW, completing allocation entirely in hardware. If the library declines or cannot allocate, the request falls back to the OS page-fault handler. This design achieves hardware-class latency and energy efficiency while supporting policies that adapt to real-time microarchitectural conditions.
\begin{figure*}[b]
    \centering
    \includegraphics[width=1.0\textwidth]{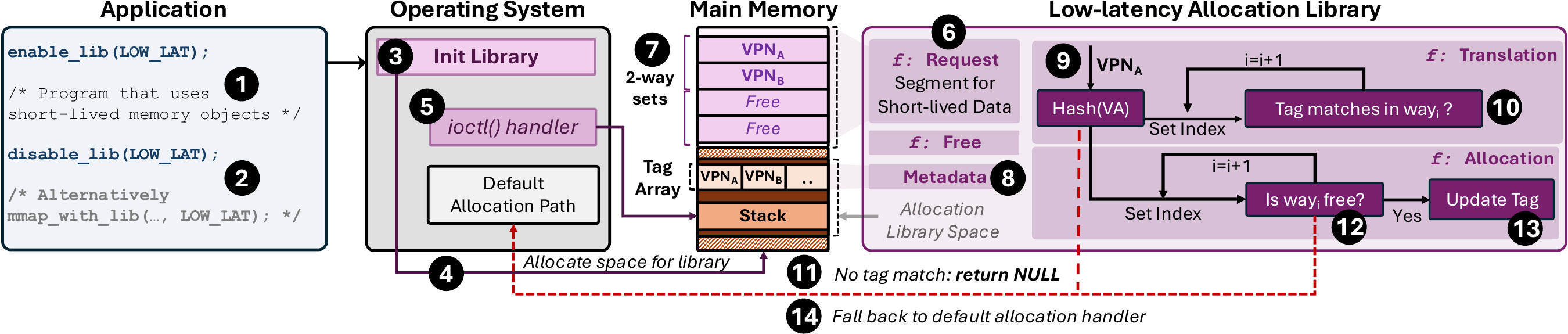}
    \caption{Valinor's software architecture (Low-latency allocation library example).
    }
    \label{fig:valinor_software}
\end{figure*}

\section{Valinor: Detailed Design}

This section presents \textbf{Valinor}’s architecture.
We first describe the application/OS interface that lets applications bind memory regions to hardware-backed allocation policies (\S\ref{sec:design:osinterface}).
We then outline the structure of the allocation libraries that implement these policies (\S\ref{sec:design:library}).
Next, we summarize the MMU extensions that identify regions using hardware-based allocation (\S\ref{sec:design:mmu}).
Finally, we describe the architecture of the PAE (\S\ref{sec:design:engine}).

\subsection{Application/OS Interface}
\label{sec:design:osinterface}

\textbf{Expressing Requirements.} In Valinor, applications (or system libraries on their behalf) explicitly express allocation requirements. This keeps the mechanism simple and under application/OS control. In practice, higher-level runtimes, such as serverless platforms~\cite{awslambda}, microservice frameworks~\cite{kubernetes}, memory allocators (e.g., TCMalloc~\cite{tcmalloc}) can automatically insert these directives, enabling transparent use of specialized hardware-based allocation without source code changes.

\textbf{Interface.} Userspace applications specify which memory regions should use hardware-based allocation through custom libraries in two ways: enabling a library for the entire process (covering all its objects, e.g., via \texttt{mmap}), or binding a single object to a library. \bqnote{Q1}\revB{In Valinor, an allocation library is not user-space code that executes on the application core. It is trusted PAE code plus metadata selected by the OS and executed by the PAE.} Valinor provides three system calls: \texttt{enable\_lib(<LIB\_NAME>)} and \texttt{disable\_lib(<LIB\_NAME>)} for process-wide control, and \texttt{mmap\allowbreak{}\_with\_lib(..., <LIB\_NAME>)} to attach a single memory object.
As shown in Fig.~\ref{fig:valinor_software}, an application can wrap a code region with \texttt{enable\_lib}/\texttt{disable\_lib} to indicate that it allocates short-lived objects and should use a low-latency library~\circled{1},\circled{2}.
Upon receiving this directive, the OS (i) allocates space for the library’s code and metadata~\circled{3}, (ii) associates the process or VMA with the library ID~\circled{4}, and (iii) programs a Control and Status Register (CSR) to inform the allocation engine about the library’s code and data locations.

\begin{figure*}[b]
    \centering
    \includegraphics[width=1.0\textwidth]{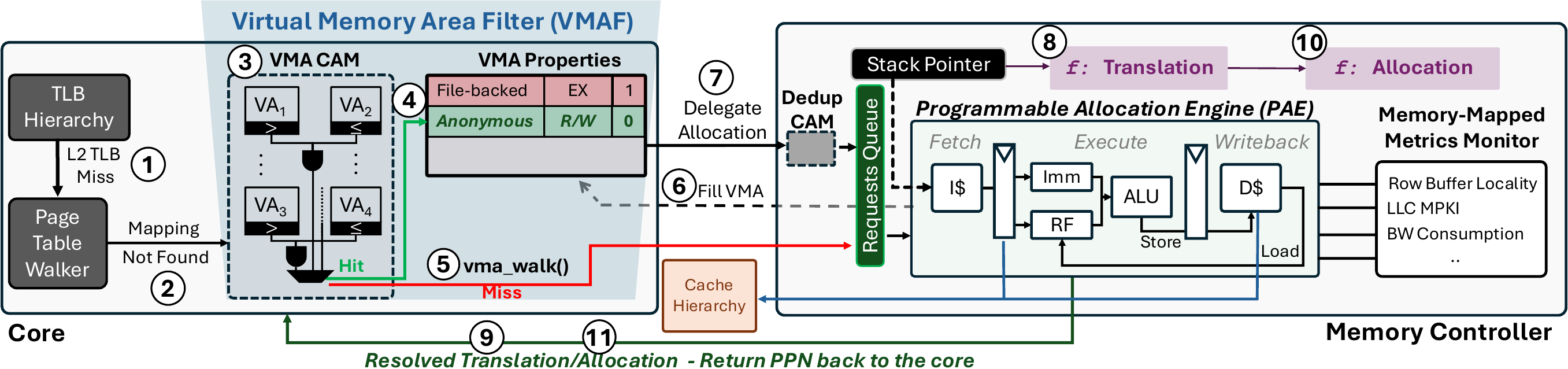}
    \caption{Valinor's hardware architecture.
    }

    \label{fig:valinor_hardware}
\end{figure*}

\subsection{Allocation Library Design}
\label{sec:design:library}

\textbf{Library Structure.} Valinor relies on custom libraries that guide the PAE's operation. 
To enable a complete memory management solution, every library needs to implement specific functions, as well as specify to the OS how to initialize/configure it.
\bqnote{Q1}\revB{Table~\ref{tab:library_functions} also lists who invokes each function: the application only selects policies through system calls, whereas the PAE executes the routines and the OS services privileged operations, such as segment grants.}
These functions are summarized in Table~\ref{tab:library_functions}.

\cqnote{Q1}\textbf{Example: Low-latency Allocation.} In Fig.~\ref{fig:valinor_software}, we illustrate an example of a library optimized for low-latency allocation via set-associative page placement.

\begin{table}[t]
\centering
\footnotesize
\caption{Memory Allocation Library Functions}
\label{tab:library_functions}
\renewcommand{\arraystretch}{1.15}
\setlength{\tabcolsep}{3pt}
\begin{tabular}{|m{2.15cm}|m{1.55cm}|m{4.35cm}|}
\hline
\textbf{Function} & \textbf{Caller} & \textbf{Description} \\
\hline
\texttt{ioctl()} &
\revB{OS} &
Used for library initialization. \\
\hline
\texttt{request(type, size)} &
\revB{PAE/OS} &
Requests a memory segment of size \texttt{size} from the OS (or releases one to it).
Grants the library ownership of a physical memory region that it will manage autonomously.
The \texttt{type} parameter specifies the operation (e.g., acquire or release). \\
\hline
\texttt{alloc(pid, va, prot)} &
\revB{PAE} &
Allocate memory for \texttt{va} and permissions \texttt{prot} for process with \texttt{pid} \\
\hline
\texttt{translate(pid, va)} &
\revB{PAE} &
Determine whether \texttt{va} has been mapped autonomously by the PAE
for process \texttt{pid}.
If so, return the corresponding PPN and permissions. \\
\hline
\texttt{free(pid, $va_1$, $va_2$)} &
\revB{OS} &
Delete all mappings for pages with virtual addresses within the range $[va_1, va_2]$ for process with \texttt{pid}.
The ownership of the freed pages remains with the library. \\
\hline
\texttt{vma\_walk(pid, va)} &
\revB{PAE} &
Locate the VMA containing \texttt{va} for process \texttt{pid} and return its properties (permissions, type).
\textbf{OS provides a default implementation} matching the kernel's VMA layout. Library writers may override it to support alternative layouts. \\
\hline
\end{tabular}
\end{table}

\dqnote{Q1}\noindent\textit{\textbf{Configuration:}} At startup, the library requests initialization information from the OS via an \texttt{ioctl}~\circled{5}. The handler and specification are provided by the library writer and may include parameters such as the size of the autonomously-managed segment or the root of a process’s page table.

\noindent\textit{\textbf{Request:}} On its first handled fault, the low-latency library requests a memory segment from the OS to use as its privately managed region, avoiding trapping into the OS upon each allocation/free~\circled{6}.
The library issues a \texttt{request} call with a \texttt{type=request} and the desired \texttt{size}. The OS allocates the segment and returns its starting physical address.
The library then organizes the segment as a set-associative structure to simplify metadata management: it divides the region into multiple sets, each with a fixed number of pages, and initializes a tag array to track free/allocated pages.

\noindent\textit{\textbf{Translation: }} When an allocation request is received from the PAE, the library first checks whether the requested virtual address (VA) has already been mapped~\circled{7}.
To achieve this, it uses a simple hash function on the VA to determine the set to look into~\circled{9}, and then scans the tags of the pages in the set to check for a hit.
If a match is found, the library returns the corresponding PPN and permissions~\circled{10}.

\noindent\textit{\textbf{Allocation: }} If the translation function does not find a mapping for the requested VA, the library proceeds to allocate a new page for it.
To do so, it uses a hash function on the VA to determine the set to allocate from~\circled{9}.
The library then scans the tags of the pages in the set to find a free page.
If a free page is found, the library updates its tag to indicate that it is now allocated and returns the corresponding PPN and permissions to the PAE~\circled{10}.
If no free page is found, the library, the allocation fails, and the library returns~\textit{NULL} to indicate that the fault needs to be handled by the OS~\circled{11}.
\bqnote{Q3}\cqnote{Q1}\revB{Thus, a set conflict in the low-latency library is a performance miss, not a correctness issue: if the selected set is full even though another set has free pages, the library declines and the normal OS allocator handles that page. Other libraries can use higher associativity, multiple hashes, or global free lists to reduce such misses.}
\\
\vspace{-1mm}
\noindent\textit{\textbf{Free: }} When the OS invokes the \texttt{free} function,
the library scans through its tag array to identify and invalidate any mappings that fall within the specified range of VAs~\circled{12} and zeroes out the pages in its data segment corresponding to the freed mappings~\circled{13}. In this way, the library reclaims the pages for future allocations without releasing them back to the OS.

\subsection{Hardware Support}
\label{sec:design:mmu}

We will now describe (i) how the MMU discovers whether an allocation should be handled by the PAE, and (ii) the architecture of the PAE. 
Fig.~\ref{fig:valinor_hardware} illustrates the key components of Valinor's hardware design and flow of operations.

\subsubsection{\textbf{Discovering Memory Object Properties}}
\label{sec:vma_lookup}

Upon a TLB miss, the MMU invokes the PTW to resolve the virtual-to-physical translation~\circledwhite{1}.
If the PTW process fails to find a valid mapping~\circledwhite{2}, it needs to determine whether the requested memory object is eligible for hardware-based allocation using the PAE.
To do so, the PTW needs to discover the properties of the memory object being accessed (e.g., file-backed or anonymous, and its access permissions).
The reason for this is twofold: (i) the PAE may be configured to handle only specific types of faults (e.g., minor faults for anonymous pages), and (ii) the PAE's functions need the access permissions of the memory object being accessed.
All this information is maintained by the OS in its VMA structures~\cite{vma}, which are kernel-specific (e.g., red-black tree~\cite{vma} or maple tree~\cite{linux-613}).
Each process maintains a list of VMAs that cover its entire VAS.

\textbf{VMA Filter.} To obtain a memory object's properties, the PTW must (i) locate the VMA covering the virtual address and (ii) extract its attributes. Valinor adds a hardware VMA Filter (VMAF) to perform these steps efficiently.
The VMAF has two components: a small CAM that caches recently accessed VMA entries (VMA CAM) and a VMA Properties table that stores the attributes of cached VMAs.
The VMA CAM holds several entries (16 in our design), each storing the bounds of a VMA and comparators to check whether an address lies within them.
The $i$-th entry of the VMA Properties table holds the metadata for the $i$-th CAM entry.

\textbf{Lookup Flow.} When the VMAF receives a request for a particular VA from the PTW, it first looks the address up in the VMA CAM~\circledwhite{3}. If the entry lies within the bounds of a CAM entry, then the corresponding properties are immediately fetched from the VMA Properties table~\circledwhite{4}. Otherwise (on a CAM miss~\circledwhite{5}), the PAE executes the library's \texttt{vma\_walk()} routine to traverse the OS's VMA structures.
If no entry is found, then the requested VA is invalid. Otherwise, the routine returns the relevant VMA information, which is placed in an entry of the VMA CAM and Properties table, evicting an existing entry if necessary~\circledwhite{6}. Finally, the properties are returned to the PTW.

\subsubsection{\textbf{Programmable Allocation Engine (PAE)}}
\label{sec:design:engine}
A lightweight compute unit implemented as a simple three-stage in-order pipeline with small (4\,KB each) instruction and data caches~\cite{riscv_mini}.
The caches are backed up by the last-level cache (LLC) to provide low-latency access to larger data structures, instead of fetching them from main memory.
To avoid the cross-core coherence, the PAE is placed inside the memory controller rather than within per-core MMUs.%
\bqnote{Q1}\revB{The PAE executes trusted allocation-library code and accesses library-owned metadata. It does not require the application core to execute or directly modify this code.}

\noindent{\textbf{ISA}.}
\aqnote{Q3}Allocation libraries are executed using a compact \texttt{RISC-V RV32I}-compatible ISA~\cite{riscv} that supports arithmetic, logical, memory, 
and control-flow operations. We extend this ISA with custom instructions for reading microarchitectural telemetry 
(e.g., per-bank DRAM conflict counters), enabling us to implement adaptive, telemetry-driven allocation policies (see~\S\ref{sec:design:specialized_libraries}).

\noindent{\textbf{Operation Flow}.}
VMAF properties are used to check hardware-based allocation eligibility of the memory object.
If the object is not eligible, the PTW directly falls back to the OS page-fault handler.
Otherwise, it forwards the request and parameters (e.g., PID) to the PAE~\circledwhite{7}.
A deduplication CAM filters out the request, in case there exists an in-flight one which already target the same VA. 
Otherwise, the PAE runs the library’s \texttt{translate} function to determine whether the VA is already mapped~\circledwhite{8}.
If a mapping exists, it returns the PPN and permissions to the PTW~\circledwhite{9}, which updates the MMU and resumes execution.
Otherwise, the PAE executes the library’s \texttt{alloc} function~\circledwhite{10}.
On success, it returns the new mapping~\circledwhite{11}. On failure, it returns \textit{NULL}, indicating the OS must handle the fault.
\bqnote{Q5}\revB{Valinor intentionally does not install every PAE-created mapping into the OS page table on this fast path. Instead, the mapping is cached in the TLB and recorded in the library metadata. If the TLB later evicts the entry, the PTW may miss in the page table and invoke the PAE's \texttt{translate} routine. This is still a hardware-executed lookup over compact metadata rather than a software page-fault handler. When the OS needs page-table visibility, it can install these mappings through the drain/revoke path described in \S\ref{sec:design:integration}.}

\subsection{System Integration and Multi-Tenant Deployment}
\label{sec:design:integration}

\textbf{Library Selection.}
Each process may enable one or more allocation libraries. The OS maintains a per-process (or per-VMA) association
between memory regions and library identifiers. During a TLB miss, the PTW consults the VMAF to
determine which library (if any) governs the faulting region. The selected library's code and data
segments are specified to the PAE via per-library CSR pointers. This indirection allows the PAE
to switch between libraries on a per-fault basis, without flushing internal state or performing
re-initialization.

\textbf{Library Sharing.}
Valinor allows multiple processes to share the same allocation library.
The library’s code is loaded once, and the PAE invokes it on behalf of any process that has bound a region to that library.
During execution, the PAE passes the caller’s PID (and other request context) to the library’s routines (e.g., \texttt{alloc}). Using this PID, the library is responsible for ensuring correctness. For example, it distinguishes requests from different processes and manages their memory independently.

\textbf{Coordination Across Multiple PAEs.}
In systems with multiple memory controllers, each controller hosts its own PAE.
Because the physical frame is determined by the library's \texttt{alloc} routine, library binding cannot depend on the physical address.
Instead, Valinor binds each allocation library instance to a specific PAE at configuration time. When the OS loads a library, it selects a target PAE (e.g., round-robin or load-balanced) and records this binding. 
On a TLB miss, the PTW identifies the library for the faulting region and forwards the request directly to its assigned PAE.
This keeps all library-related state local to one PAE, avoiding inter-PAE coordination, while allowing the OS to flexibly place libraries and tune performance without modifying the MMU or PTW.
\commonqnote{CQ2}\revCQ{This binding is per library instance rather than a fundamental one-policy/one-channel limit. The OS can instantiate the same allocation policy on multiple PAEs and bind different VMAs, or OS-created subranges of a larger mapping, to different instances. This preserves local metadata per PAE while allowing channel striping or load balancing at VMA granularity. Tier-aware policies can be expressed similarly. The OS grants a library instance a segment from the desired tier (e.g., local DRAM, remote NUMA memory, or CXL memory). Dynamic tier migration first drains or revokes the affected segment so that the standard OS migration machinery can operate on installed PTEs.}

\textbf{Supporting fork / Copy-on-Write (CoW) semantics.}
Because Valinor maintains VA-to-PA mappings in both the OS page table and the library's tag array, explicit synchronization is needed before \texttt{fork()} to preserve CoW semantics.
Valinor addresses this with a lightweight \emph{drain} step. Before forking, the OS walks the library's tag array of the calling process, installs a standard PTE for each valid mapping, and reclaims the library's physical segment.
After draining, \texttt{fork()} proceeds using Linux's unmodified CoW machinery. The kernel marks shared pages read-only, and subsequent writes trigger standard CoW faults, handled entirely by the OS.
After fork, the library is back in the initial state: its tag array is empty and it owns no segment.
Correctness follows directly from Linux's \texttt{fork()}/CoW implementation, which operates over the page table. After the drain, all mappings reside in the page table, so the existing semantics apply.
The required kernel changes are minimal: only a drain hook
is inserted before \texttt{fork()}. The \texttt{fork()} code itself and the CoW page-fault handler remain unmodified.
Since \texttt{fork()} already traverses the full VMA tree, the incremental cost of walking the compact tag array is insignificant.

When it comes to VMAF, the properties table can hold stale entries if the OS mutates VMAs
concurrently (e.g., via \texttt{mprotect()}).
Valinor exposes a programmable \texttt{vma\_walk} function (Table~\ref{tab:library_functions})
so that VMA lookup is portable across kernel VMA layouts without hardware changes.
The fenced protocol that prevents a concurrent \texttt{alloc} from committing against stale
VMA properties is defined formally as transition~(3) in \S\ref{sec:ownership}.

\aqnote{Q5}\revA{
\textbf{Portability to Non-Linux OSes.}
Although our prototype targets Linux, Valinor's design is not Linux-specific: it relies only on features common to all modern OSes (e.g., FreeBSD, Windows, and macOS) such as demand paging, \texttt{fork()}-style process creation, and page-table-based virtual-to-physical mappings.
The only OS-dependent step is walking the VMA structure, which varies across kernels.
Valinor confines this dependency to a single reprogrammable \texttt{vma\_walk()} routine, which is described in Table~\ref{tab:library_functions}.
}

\textbf{Security Analysis.}
We assess the attack surface introduced by Valinor's architectural extensions across three threat categories.
Our attacker model assumes that the the PAE library code is trusted by the kernel.
\aqnote{Q4}\dqnote{Q3}\revA{This assumption is justified because user-space cannot load libraries or modify the PAE's code: as detailed in \S\ref{sec:design:osinterface}, applications only express preferences (e.g., \texttt{enable\_lib}), while the OS retains authority over which library is deployed and over the \texttt{CSR}s configuring it.
Because the PAE executes the code referenced by these OS-managed \texttt{CSR}s, a malicious process can neither inject code nor hijack the PAE's control flow.
Thus, the attacker model trusts the OS-loaded library code.
Valinor adds no new software privilege-escalation vectors, inheriting the OS's existing isolation and permission boundaries.}

\aqnote{Q1}\revA{\textbf{Stack corruption \& privilege escalation.} Since the library is trusted and assumed correct, a library-bug stack corruption is out of scope. Even a faulty library exposes no stack-corruption surface beyond that of a baseline Linux system.
Moreover, libraries are installed by the kernel: they are loaded into main memory and bound via OS-managed \texttt{CSR}s. Thus, libraries execute within the kernel's trust domain, rather than under user control, so a user-space process cannot corrupt library code to escalate to kernel privileges.}

\commonqnote{CQ1/CQ2}\revCQ{\subsection{Ownership Model and Coordination Protocol}
\label{sec:ownership}

Valinor coordinates the OS and the PAE(s) through a single \emph{exclusive
ownership} invariant rather than general hardware cache coherence.

\emph{For every virtual page, at most one owner (the OS or exactly one PAE) may
install or modify its mapping at any instant. Ownership changes only through an
explicit, serialized handoff that carries a writeback/invalidate barrier.}

If this invariant holds, conflicting mappings are impossible by construction. A
race requires two concurrent writers to the same mapping, which the invariant
forbids. Below we define the ownership states, the transitions between
them, and the rules that preserve the invariant under concurrency. The submitted
design already enforces the invariant on the common paths. The revision makes the
two barrier rules below explicit to close the remaining concurrency corner
cases.

\noindent\textbf{Ownership states.} A region is \emph{OS-owned} by default. In
this state, the OS handler installs mappings and the page table is authoritative.
A region can also be \emph{PAE-owned}
(marked \texttt{VM\_HW\_ALLOC} and bound to one PAE, whose metadata is
authoritative and for which the OS handler skips allocation). Finally, a region
can be \emph{in-transit} while ownership is being handed off. No new mapping may
be installed for the affected addresses until the handoff completes.

\noindent\textbf{Transitions.}
\emph{(1) Bind (OS$\rightarrow$PAE):} the OS allocates the library's code and
metadata, sets \texttt{VM\_HW\_ALLOC}, and programs the PAE CSRs.
\emph{(2) Drain/revoke (PAE$\rightarrow$OS):} used by \texttt{fork},
\texttt{munmap}, reclaim, and migration. It is defined as a barrier below.
\emph{(3) VMA mutation (in place):} for changes that do not transfer ownership
(e.g., \texttt{mprotect}), defined as a fenced critical section below.

\noindent\textbf{Drain as a barrier (transition~2).} To guarantee that
a fault racing with drain cannot escape the barrier, the revision adds a
request-queue fence to the drain path. On drain initiation for a region, the PAE
\emph{(i)} stops admitting new requests whose VPN falls in the region,
\emph{(ii)} completes all in-flight requests for the region. Each either
installs into the library metadata or declines. The PAE then \emph{(iii)} writes
back its metadata, and only then \emph{(iv)} signals drain-complete. The OS
resumes authority strictly after step~(iv). A fault arriving after
step~(i) is held until drain-complete and then resolved by the OS on the
now-installed page-table entries. Every fault is thus ordered either entirely before
the snapshot (drained into a PTE) or entirely after it (handled by the OS). None
straddles the handoff. The step~(iii) writeback is also the barrier that makes
CPU/PAE cache coherence unnecessary.

\noindent\textbf{VMA mutation as a fenced critical section (transition~3).}
Serializing \texttt{vma\_walk} against VMA mutation, as in the submitted design,
prevents the VMAF from observing a torn update but does not by itself prevent an
\texttt{alloc} that has already read a region's VMA properties from committing a
mapping against properties the OS subsequently changes (e.g., an \texttt{mprotect}
that revokes write permission). The revision closes this case by treating a VMA
mutation as a small drain of the affected region. Under the CSR lock, the OS
\emph{(i)} marks the affected VMAF entries stale and stops new PAE requests to the
region, \emph{(ii)} drains in-flight PAE requests touching that VMA to completion,
\emph{(iii)} updates the VMA tree and invalidates the affected VMAF CAM entries
(analogous to a TLB shootdown), and \emph{(iv)} releases the lock. No
\texttt{alloc} can therefore commit against stale properties. Because VMA
mutations are far rarer than faults, this fence is off the common-case path.

\noindent\textbf{Intra-state serialization (same-VA faults).} Concurrent faults
on the same VA are \emph{not} an ownership change and are resolved within the
current owner. A VPN-indexed deduplication CAM at the PAE request queue admits a
single in-flight resolution per VA. The entry persists until resolution
completes, including any OS fallback. If the PAE declines, it sets
\texttt{PAE\_DECLINED}, authorizing the OS to handle that one fault without
violating single ownership. Two threads faulting on the same VA therefore cannot
produce conflicting mappings.

\noindent\textbf{Coherence as a corollary.} Because the invariant guarantees no
two agents ever hold write rights to the same mapping or metadata
simultaneously, there is no concurrent-writer scenario for hardware coherence to
resolve. PAE caches hold only owner-private data. The sole
OS$\leftrightarrow$PAE-shared structure, VMA state, is kept consistent in
software by the fenced transition~(3) described above. Coherence is thus
replaced by explicit, barrier-carrying ownership handoff. This is the same
discipline used by device DMA and accelerator scratchpads.
}

\definecolor{lightgray}{gray}{0.95}
\definecolor{darkblue}{RGB}{20,20,120}
\definecolor{darkgreen}{RGB}{0,100,0}

\lstdefinestyle{valinor}{
  backgroundcolor=\color{lightgray},
  basicstyle=\ttfamily\footnotesize,
  keywordstyle=\color{darkblue}\bfseries,
  commentstyle=\color{darkgreen}\itshape,
  numbers=left,
  numberstyle=\tiny\color{gray},
  numbersep=5pt,
  frame=single,
  frameround=tttt,
  rulecolor=\color{gray},
  breaklines=true,
  tabsize=2,
  showstringspaces=false,
  captionpos=b,
  aboveskip=6pt,
  belowskip=6pt,
}

\section{Specialized Allocation Libraries}
\label{sec:design:specialized_libraries}

\cqnote{Q1}Valinor supports a wide range of specialized allocation policies besides the low-latency one. 
These policies are not intended to be the optimal or most sophisticated 
designs for their respective goals. They are simple, representative policies 
constructed and evaluated to show how diverse 
allocation strategies can be realized using Valinor’s programmable substrate. 
Below, we present three such libraries and evaluate them in \S\ref{sec:results:sim_interference}. We additionally describe a \commonqnote{CQ3}\revCQ{fourth library} that illustrates tier-aware placement (\S\ref{sec:tier-aware}).

\bqnote{Q4}\subsection{\revB{Translation-Metadata Integrity} Library}
\label{sec:design:specialized_libraries:integrity}

\revB{\textbf{Threat model.} This library addresses a specific threat. An attacker may corrupt PAE-managed mapping metadata (e.g., a VA$\to$PA entry stored in the library\textquotesingle s data segment) and cause the PAE\textquotesingle s \texttt{translate()} to return a wrong PPN. It does \emph{not} protect the data stored inside allocated pages. A data write does not change the VA$\to$PA mapping, so no MAC or Merkle update is triggered by application stores. This check also hardens the PAE-metadata attack surface identified in \S\ref{sec:design:integration}. The PAE\textquotesingle s request queue can leak allocation patterns. Authenticating the metadata entries prevents an attacker from using such observations to inject a forged mapping.}

\revB{This library augments the low-latency allocation and translation path of
\S\ref{sec:design:library} with lightweight integrity validation. Its goal is to protect
the integrity of the \emph{translation metadata}. These are the VA-to-PA mappings the PAE
stores and returns, not the data held in the allocated pages. The threat
it addresses is corruption of the library's mapping metadata in DRAM, whether
from a hardware fault, a memory-safety bug, or an adversary wielding a
DRAM-tampering primitive such as Rowhammer~\cite{citadel}. Because the PAE has
OS-level authority to return a PPN directly to the MMU, a single silently flipped
PPN in the metadata would cause \texttt{translate} to hand a process a physical
frame it was never granted. This creates an illegal remapping attack against the
trusted PAE itself. The integrity check ensures such a corrupted mapping is
detected and rejected before it is installed.

We implement two variants. The \emph{per-page MAC} variant stores a 64-bit MAC
alongside each metadata entry in the library's data segment. On a \texttt{translate}
request, the library recomputes the MAC over the tuple \texttt{(pid, va, prot,
PPN)} and a secret key, and compares it against the stored value. The \emph{Merkle
tree} variant instead builds a tree over the mapping metadata and persists only
the root hash~\cite{merkle1987digital}. Validation checks the queried entry against its
authentication path up to the stored root, trading more hashes per lookup for
constant-size persistent state and detection of any tampering anywhere in the
metadata. The secret key and the Merkle root are held in PAE-private state (a CSR
and the PAE's protected metadata region, respectively). They are never stored in
the library-owned DRAM segment that the threat model treats as corruptible. Thus,
an attacker who can flip DRAM bits cannot forge a matching MAC or root.

Authentication state is created or updated only when the library installs,
removes, or remaps an entry, never on ordinary writes to a page's data, since
those do not change the mapping. This is what keeps the mechanism off the
common-case path: a \texttt{translate} adds one MAC verification (or one
authentication-path check), and \texttt{alloc} adds one MAC/root update. On any
validation failure the library returns \texttt{NULL}, the PTW falls back to the
OS, and the OS is notified of an integrity violation. The library is therefore a
self-defense mechanism for the PAE's own metadata and is orthogonal to full
data-at-rest integrity, which a separate memory-integrity engine can provide by
authenticating data writes. To keep a protected region fully covered, the library
issues a \texttt{request} to the OS for an integrity-managed segment whenever an
allocation would otherwise spill outside its current segments.}

\subsection{Speculation-Enhanced Allocation Library}
\label{sec:speculation}
\label{sec:design:specialized_libraries:spec}

This library accelerates common sequential and strided allocation patterns by
predicting and pre-establishing future page mappings, extending the low-latency
allocator of \S\ref{sec:design:library} with two lightweight structures: a stride
detector that tracks the deltas between successive faulting VAs per region, and a
small speculative buffer that holds pre-allocated mappings not yet claimed by a
fault. When the detector observes a stable stride across consecutive allocations
(e.g., the linear page-touch pattern of a JSON parser streaming through a buffer),
the library pre-allocates the next few pages from its private physical segment and
records them in the speculative buffer, performing this work \emph{off} the demand
path while the PAE would otherwise be idle.

The value of speculation here is latency hiding rather than throughput. When a
subsequent fault matches a buffered prediction, the library commits the
pre-allocated page immediately. It returns the PPN with no allocation work on the
critical path, so the fault is resolved in the time of a metadata lookup. This is
why, in our evaluation (\S\ref{sec:results:sim_interference}), the speculation library hides over 99\%
of page-fault latency even on the simple in-order PAE. The actual allocation has
already happened before the fault arrives, leaving only commit on the hot path.

Speculation must not violate the single-ownership invariant of
\S\ref{sec:ownership}, so a speculatively pre-allocated page is owned by the
library but \emph{unmapped} until a matching fault commits it. It is never
installed in the TLB or page table on speculation alone. The deduplication
CAM still admits only one in-flight resolution per VA, so a real fault and a
speculative pre-allocation for the same VA cannot both commit. On a misprediction
or stride change, the library simply reclaims the unclaimed buffered pages back
into its segment's free pool. No OS interaction and no rollback of installed
state are required, because nothing was installed. If no prediction matches, the
library falls back to ordinary allocation, or, if its segment is exhausted, lets
the PTW invoke the OS. Freeing a page also clears any speculative state derived
from it. The result preserves the low-latency critical path while opportunistically
eliminating allocation cost for predictable workloads, at the cost of a small
speculative buffer and the pages held in it.

\commonqnote{CQ3}\revCQ{\subsection{Tier-Aware Placement Allocation Library}
\label{sec:tier-aware}

This library steers allocations across heterogeneous memory tiers (e.g., local
DRAM, remote NUMA memory, and CXL-attached memory) according to per-region intent
and runtime telemetry. This demonstrates that tier-aware placement, which
fixed-function allocators cannot express, is realizable within Valinor's
segment-ownership model without cross-PAE coordination or fast-path relocation.
Unlike the single-segment libraries of \S\ref{sec:design:library}, it is granted one
physical segment per participating tier at configuration time: the OS issues a
\texttt{request} per tier and tags each returned segment with its tier identity
and cost profile (access latency, bandwidth budget, and capacity). Each segment
is organized as an independent per-tier free pool. A single global
open-addressing hash table holds the VA-to-PA mapping, so a \texttt{translate}
remains one lookup on the critical path and never needs to know which tier served
a page.

On an \texttt{alloc} request, the library scores every candidate tier with a
single continuous utility objective, ranks the tiers best-first, and places the
page in the highest-scoring tier that is not conflicted (full or over budget):
\begin{equation*}
\mathrm{utility}(t) = V(\mathrm{hint}) \cdot \mathrm{locality}(t)
  - w_{\mathrm{bw}} \cdot \mathrm{bw\_util}(t)
  - w_{\mathrm{cap}} \cdot \mathrm{cap\_util}(t).
\end{equation*}
The locality term $\mathrm{locality}(t) = (\mathit{lat}_{\max} -
\mathit{access\_ns}(t)) / \mathit{range}$ is $1$ for the fastest tier and $0$ for
the slowest. The per-VMA intent hint sets its sign and weight. $V =
+\mathit{locality\_value\_hot}$ for latency-sensitive regions pulls allocations
toward fast tiers, while $V = -\mathit{locality\_value\_cold}$ for
capacity-oriented regions actively pushes cold data off fast memory. The two
penalty terms consume live telemetry read through the PAE's custom ISA extensions
(\S\ref{sec:design:engine}). $\mathrm{bw\_util}(t) = \mathit{recent\_alloc\_count} /
\mathit{bw\_capacity}$ is windowed bandwidth demand, so tiers nearing their
bandwidth budget are penalized, and $\mathrm{cap\_util}(t) = \mathit{occupied} /
\mathit{capacity}$ penalizes near-full tiers, with a full tier scoring $-\infty$
and being excluded outright. This single objective generalizes cleanly to $N$
tiers, subsuming the simpler binary local-versus-far spill as a special case.
\texttt{free} returns each page to its originating tier's pool, which the
per-mapping tier tag makes unambiguous.

Crucially, the library performs tier-aware \emph{initial placement} entirely on
the hardware fast path but never relocates a page after allocation. Dynamic
migration remains an OS responsibility carried out through the standard drain/revoke path
(\S\ref{sec:design:integration}). This includes promoting a region that turns hot or
demoting a cold one. The OS drains the PAE, installs the affected
mappings as ordinary PTEs, and applies its existing migration machinery (e.g.,
AutoNUMA). This division is what lets Valinor express tiering that prior
fixed-function allocators fundamentally cannot. A greedy hardware allocator is
tier-blind and scatters latency-sensitive data into slow memory, whereas the
programmable PAE makes a telemetry- and intent-aware decision per allocation. The
latency-critical path still stays in hardware, and the OS retains full authority over
relocation and reclamation.}

\section{Evaluation Methodology}
\label{sec:methodology}

\vspace{-1mm}
\definecolor{SoftPeach}{rgb}{0.937,0.901,0.901}
\begin{table}[h!]
\centering
\scriptsize
\caption{Simulation Configuration and Systems}
\vspace{-2mm}
\label{tab:simconfig}
\begin{tblr}{
  width = \linewidth,
  cell{1}{1} = {c=2}{0.94\linewidth},
  cell{12}{1} = {c=2}{0.94\linewidth},
  cell{3}{1} = {r=4}{},
  cell{7}{1} = {r=2}{},
  cell{9}{1} = {r=2}{},
  colspec = {Q[170]Q[600]},
  row{1,12} = {SoftPeach,c},
  vlines, hlines
}
\textbf{RISCV Prototype Configuration} & \\
\textbf{Core} & 64-bit RISC-V BOOM Out-of-Order core. 300MHz on FPGA\\
\textbf{MMU} & L1 I-TLB: 32-entry, direct-mapped, 1-cycle latency\\
 & L1 D-TLB (4 KB): 8-entry, fully assoc, 1-cycle latency \\ 
 & L2 TLB: 512-entry, direct-mapped, 2-cycle latency\\
 & Page Walk Cache: 8-entry, fully assoc, 1-cycle latency\\
\textbf{Cache} & {L1 I/D-Cache: 32 KB, 4-way assoc, 3-cycle access latency}\\
 & LRU replacement policy\\
\textbf{Allocation \newline Engine} & {Core: RISCV-MINI 32-bit in-order core \cite{riscv_mini}. 300MHz on FPGA} \\
& L1 I/D-Cache: 4KB, direct-mapped , 1-cycle latency \\
\textbf{FPGA} & {Xilinx ZCU 106 \cite{zcu106}, 1GB DDR4 \textbf{Linux Kernel}: v5.11.6} \\
\hline
\textbf{Virtuoso+Sniper Configuration} & \\

\textbf{Core} & 4-way OoO x86, 2.9 GHz \\

\textbf{MMU} & L1 ITLB: 128 entries. L1 DTLB (4KB): 64 entries.
                 L1 DTLB (2MB): 32 entries. L2 TLB: 2048 entries.
                 3 PWCs, 32 entries \\
\textbf{Caches} & L1 I/D: 32 KB. L2: 2 MB. L3: 2 MB/core \\
\textbf{DRAM} & 256 GB DDR4-2400 \\
\hline
\textbf{Real CPU} & Linux 5.15. 256 GB DDR4-2400. Intel Xeon Gold 6226R \\
\end{tblr}
\end{table}

Table \ref{tab:simconfig} summarizes the configuration of the real RISC-V prototype, the simulated system used in our evaluation and the real server-grade CPU used for the experiments shown in \S\ref{sec:motivation}.

\textbf{RISC-V Soft-core Prototype}
We prototyped our proposed design and tested it on a Xilinx ZCU106 FPGA~\cite{zcu106} to (i)~showcase the feasibility of implementation of our key mechanism and
(ii)~evaluate the performance benefits of Valinor on a real system running a full-blown OS. 
To model the PAE, we employed a small 32-bit in-order RISC-V core~\cite{riscv_mini}.
We integrated this unit within the translation subsystem of the BOOM OoO core to accommodate allocations in hardware. The prototype runs Linux kernel v.5.11.6~\cite{linuxkernel}.

\textbf{Simulation}
We used Sniper~\cite{sniper} in combination with Virtuoso~\cite{kanellopoulos2025virtuoso} to simulate a full-system x86 environment with detailed modeling of the MMU and memory hierarchy. 
We configured MimicOS (the lightweight OS modeled in Virtuoso) to mimic Linux's minor page-fault handling behavior.
Virtuoso was extended to model the Valinor architecture, including the PAE, allocation libraries, and their interactions with the OS and MMU.
We will open-source the RISCV prototype and our simulation infrastructure to enable further research.

\textbf{Workloads.} We use workloads from three sources. DeathStarBench~\cite{deathstarbench} is an open-source cloud microservices suite. vSwarm~\cite{vswarm} is a set of ready-to-run serverless benchmarks. Bitnet~\cite{wang20241bitaiinfra11} is an inference framework for 1-bit LLMs targeting resource-constrained devices.
From DeathStarBench, we use all hotel-reservation microservices (\texttt{hotel-res}) and a recommendation service (\texttt{rec-sys}).
From vSwarm, we use distributed compilation (\texttt{comp}), \texttt{aes}, authentication (\texttt{auth}), video analytics (\texttt{video}), compression (\texttt{zlib}), and image rotation (\texttt{img-rot}).
We also add a Python \texttt{json} parser, a Python \texttt{wcount} function, and a C++ \texttt{sqlite}-based key–value store accessed over HTTP.
From Bitnet, we use the b1.58 2B4T model (>2B parameters) and run four inference tasks using the benchmark’s generator: three with a 100-token prompt and outputs of 3, 10, and 20 tokens, and one with a 100-token prompt and a 3-token output.
All workloads run to completion in both simulation and the RISC-V prototype, with warm-up to avoid cold starts (all required binaries, libraries, images, and models are preloaded).

\section{Evaluation Results}
\label{sec:results}
\subsection{RISC-V Soft-core Prototype Results}

We first evaluate our proposed design on a RISC-V-based soft-core FPGA prototype as described in \S\ref{sec:methodology}.
We compare the performance of four different systems: i) \textbf{Baseline}: unmodified BOOM processor running unmodified Linux Kernel,
ii) \textbf{SW-only}: unmodified BOOM processor with modified Linux kernel implementing the Low-Latency Allocation library (\S\ref{sec:design:library}) solely in software,
iii) \textbf{Fixed-HW}: modified BOOM processor with fixed hardware module that implements the Low-Latency Allocation library (\S\ref{sec:design:library}) and handles allocations in hardware,
and iv) \textbf{Valinor}: our proposed design with a PAE that implements the Low-Latency Allocation library (\S\ref{sec:design:library}) and handles allocations in hardware.
We evaluate all four systems across a subset of the benchmarks described in \S\ref{sec:methodology} (the ones we could successfully run on RISC-V BOOM),
and 4 microbenchmarks that we designed to specifically stress test the allocation mechanisms. The short-lived allocation pool size is set to $16MB$ for all experiments, unless otherwise specified.

\begin{figure}[h!]
  \includegraphics[width=1.0\linewidth]{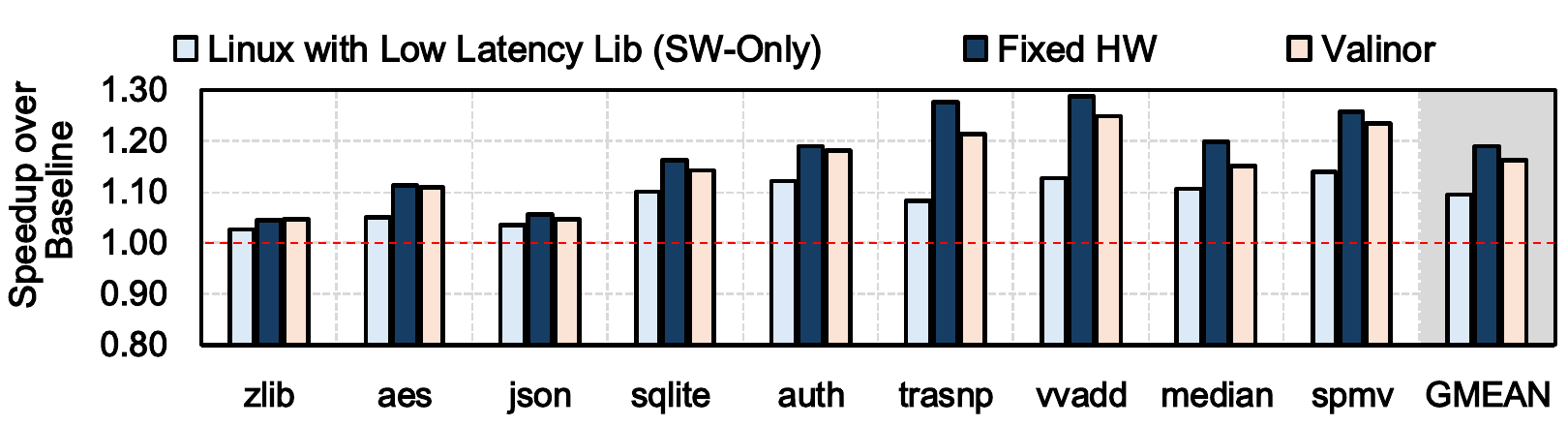}
  \caption{Performance speedup achieved by Valinor, Fixed-HW, and SW-only over Baseline Linux across various benchmarks.}
  \label{fig:all_speedups}
\end{figure}

Fig.~\ref{fig:all_speedups} shows the end-to-end performance speedup measured on the FPGA prototype for our evaluated scheme.
 We make three key observations: (i) Despite making no modifications to the HW, the \textbf{SW-only} version already leads to an average speedup of $9\%$, which is indicative of the allocation scheme's efficiency compared to the baseline Linux's page fault handler,
 (ii) the \textbf{Fixed-HW} version is the fastest, achieving an average speedup of $16\%$, thus showcasing the benefits of offloading page fault handling in HW and,
 (iii) \textbf{Valinor} consistently outperforms both the baseline and \textbf{SW-only} by 16\% and 7\% on average,  while still being only $3\%$ slower on average commpared to \textbf{Fixed-HW}. 
We conclude that, despite employing a general-purpose programmable core, \textbf{Valinor}'s performance is still comparable to that of a fixed HW module. 
Our energy measurements with Synopsys~\cite{synopsysdc} indicate that Valinor saves on average 5\% (up to 8\%) of energy over Linux (figure not shown due to space constraints).

\begin{figure}[ht!]
  \centering
  \includegraphics[width=1.0\linewidth]{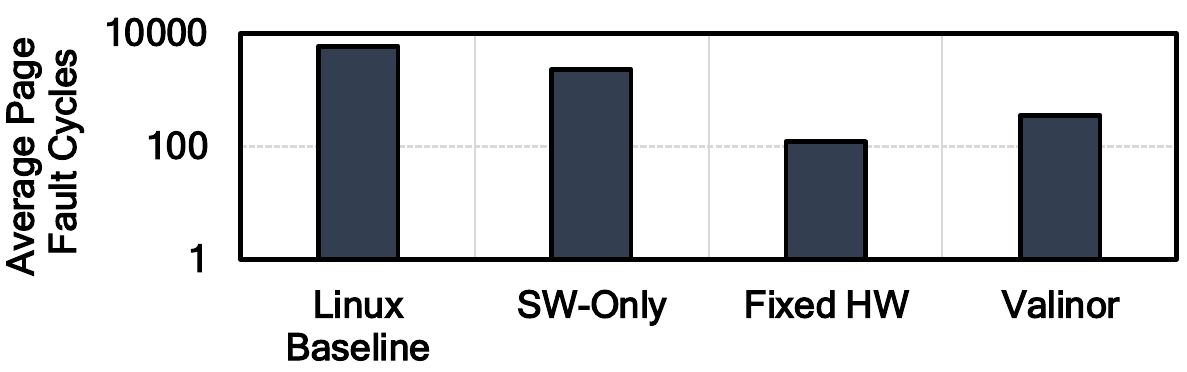}
  \caption{Average number of cycles taken per page fault for each of our evaluated designs.}
  \label{fig:avg_pf_cycles}
\end{figure}

\textbf{Page Fault Latency Analysis.}
To illustrate the performance benefits of our design, Fig.~\ref{fig:avg_pf_cycles} reports the average cycles spent on page-fault handling across systems.
We make three observations.
First, baseline Linux requires 5881 cycles per minor fault (consistent with our server-grade measurements from \S\ref{sec:motivation}).
Second, delegating allocation entirely to fixed-function HW yields a 48x speedup due to bypassing the page-fault handler. Third, although Valinor’s in-order core is ~3x slower than Fixed-HW (344 vs. 122 cycles), it still provides a 17.07x improvement over Linux.
Overall, even with a programmable fabric, moving allocation into HW offers substantial gains by avoiding the overheads of the OS fault-handling.

\begin{figure}[ht!]
  \centering
    \includegraphics[width=1.0\linewidth]{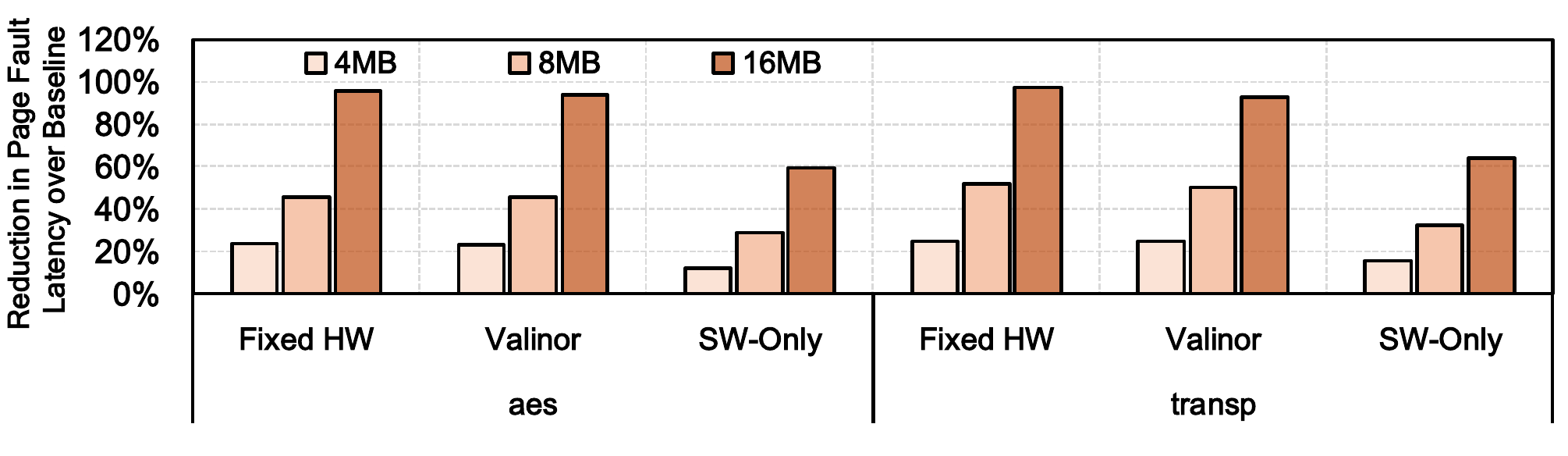}
  \caption{Page fault latency reduction across different allocation pool sizes for \texttt{aes} and \texttt{transp}.}
  \label{fig:pool_size_sensitivity}
\end{figure}

\textbf{Sensitivity Analysis on Allocation Pool Size.}
To illustrate the benefits of hardware-assisted allocation, we evaluated two benchmarks (\texttt{aes} and \texttt{transp}) across a range of allocation pool sizes (4MB to 16MB).
Fig.~\ref{fig:pool_size_sensitivity} compares the page fault latency reduction achieved by \textbf{Fixed-HW}, \textbf{Valinor}, and \textbf{SW-Only}, all relative to baseline Linux.
We make two key observations.
First, hardware methods are superior even for small pool sizes.
With a 4MB pool for \texttt{aes}, the hardware-assisted \textbf{Fixed-HW} and \textbf{Valinor} achieve latency reductions of approximately 24\% and 23\%, respectively. This is already a distinct advantage over the \textbf{SW-only} approach, which reduces latency only by 13\%.
Second, the performance gap widens drastically with larger pools.
For a pool of size 16MB, in \texttt{aes}, \textbf{Fixed-HW} and \textbf{Valinor} deliver latency reductions of 95\% and 93\%, respectively.
In contrast, the \textbf{SW-only} method's latency reduction is limited to 60\%.
The \texttt{transp} benchmark exhibits an even more accentuated trend, with \textbf{Fixed-HW} and \textbf{Valinor} achieving significantly greater reductions than the software-only solution.

\subsection{Design Space Exploration in Simulation}
\label{sec:results:sim_interference}
To complement our FPGA prototype, we evaluate Valinor in simulation as described in \S\ref{sec:methodology}.

\begin{figure}[h!]
  \centering
  \includegraphics[width=1.0\linewidth]{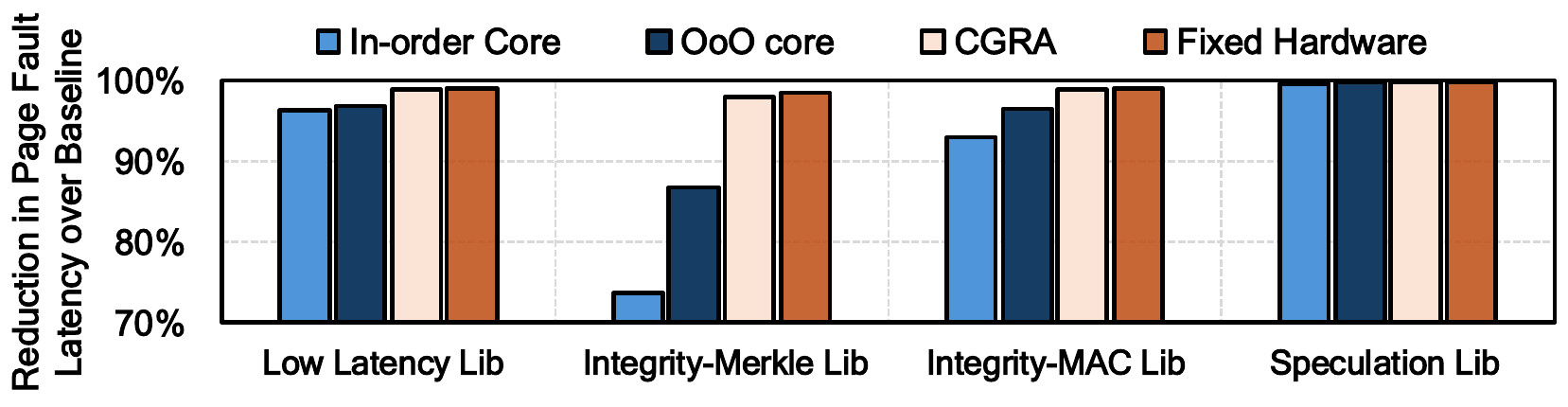}
  \caption{\revCQ{Page fault handling latency reduction across four allocation libraries and four PAE designs. Fixed HW is a fixed-function implementation of the same Valinor policy used in each group.}}
  \label{fig:design_space}
\end{figure}

\textbf{Library and Microarchitecture Exploration.}
We evaluate four allocation libraries: Low-Latency Allocation (LLA) (\S\ref{sec:design:library}), Integrity-Merkle, Integrity-MAC (\S\ref{sec:design:specialized_libraries:integrity}), and Speculation-Based Allocation (SBA) (\S\ref{sec:design:specialized_libraries:spec}).
We also prototype four PAE microarchitectures: a 3-stage in-order core, an OoO core, a CGRA similar to~\cite{hycubeDAC17}, and a fixed-function module. Libraries are ported to the CGRA using Morpher~\cite{morpher}. For fixed hardware, we use RTL-derived latencies via HLS and simulate memory accesses in Sniper~\cite{sniper}.
\commonqnote{CQ3}\revCQ{The Fixed HW bars in Fig.~\ref{fig:design_space} should be read as an upper-performance point for a single frozen policy. They are not prior greedy hardware allocators. For each bar, Fixed HW implements the same Valinor library policy being evaluated, but with that policy hardwired into a fixed-function module. Fig.~\ref{fig:design_space} therefore measures the cost of programmability for a fixed policy. It does not measure the benefit of changing policies at runtime or deploying policies that fixed hardware did not anticipate.}
Fig.~\ref{fig:design_space} shows the average page-fault latency reduction across all workloads in \S\ref{sec:methodology}, normalized to the baseline OS handler. We make four observations.
First, SBA is most effective, cutting latency by over 99\% and nearly hiding page faults even on the in-order core.
Second, for the other libraries, the in-order core still performs well: 90\% reduction for LLA, 93\% for Integrity-MAC, and 73\% for the more complex Integrity-Merkle.
Third, Integrity-Merkle is the only case where the OoO core offers a major boost, improving 73\% to 87\% by exploiting ILP.
Fourth, the CGRA and fixed hardware achieve the highest performance (96–99\%) across all libraries, though CGRAs cannot guarantee that every library maps cleanly.
Overall, Valinor’s PAE supports a wide range of libraries and delivers substantial gains even with a simple in-order core.
\revCQ{The small gap to Fixed HW is the desired result. It shows that programmability costs little when the policy is held fixed. The benefit of Valinor appears in the next two experiments, where the useful policy depends on runtime conditions or software intent. Those policies require the same silicon to execute different allocation logic. A fixed-function allocator would need a redesign.}

\begin{figure}[b!]
  \centering
  \includegraphics[width=1.0\linewidth]{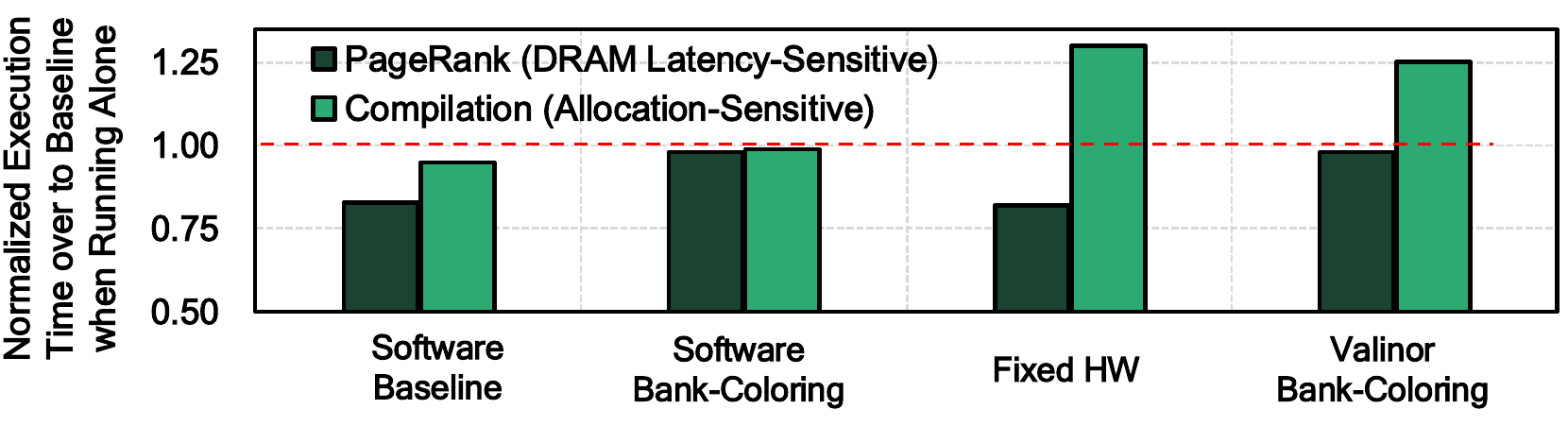}
  \caption{Normalized execution time of co-located applications. TAA (Valinor) is compared to SW-only methods and a Fixed HW allocator.}
  \label{fig:taa_results}
\end{figure}

\revCQ{\textbf{Adaptive Policies Beyond Fixed HW.}}
\revCQ{The previous experiment holds the allocation policy fixed. We now evaluate policies whose benefit comes from adapting placement decisions to runtime conditions or per-VMA intent.}

\textbf{Telemetry-driven Adaptive Allocation Library.}
To showcase Valinor's adaptability, a key benefit over fixed-function hardware, we evaluate the Telemetry-driven Adaptive Allocation (TAA) library.
We configure Valinor's PAE using the simple, in-order core design and run two applications \textbf{colocated on the same CPU}: (i) a DRAM-latency sensitive benchmark (\texttt{pagerank}) and (ii) an allocation-sensitive serverless function (\texttt{compilation}).
Our goal is to demonstrate how TAA can adapt its allocation policy based on runtime telemetry to optimize data placement and reduce DRAM bank-level interference.
Fig.~\ref{fig:taa_results} shows the end-to-end performance for both applications relative to running in isolation (the 1.0 line). We compare four scenarios: (i) the baseline OS, (ii) the OS with software bank-coloring, (iii) a Fixed HW allocator, and (iv) Valinor executing TAA.
We make three key observations.
First, in the Software Baseline, \texttt{pagerank} experiences a 17\% performance degradation (0.83x performance) due to interference from the co-located \texttt{compilation} task.
\revCQ{Second, the Fixed HW allocator in this experiment is a static low-latency allocator. It is unaware of memory-bank pressure and performs greed-for-speed allocations, so \texttt{pagerank} still suffers a ~17\% performance degradation.}
\revCQ{It still improves the allocation-sensitive \texttt{compilation} workload by removing software page-fault overhead. It cannot protect a co-located bandwidth-sensitive workload because its placement policy is fixed.}
Third, both Software Bank-Coloring and Valinor (TAA) successfully mitigate the interference on \texttt{pagerank}, restoring its performance to ~98-99\% of its baseline. However, Valinor also accelerates the \texttt{compilation} task by 1.25x, a benefit the software-only bank-coloring approach cannot provide. This demonstrates TAA's ability to adaptively manage placement to both protect victim applications and accelerate allocation-sensitive ones.

\begin{figure}[!ht]
  \centering
  \includegraphics[width=1.0\linewidth]{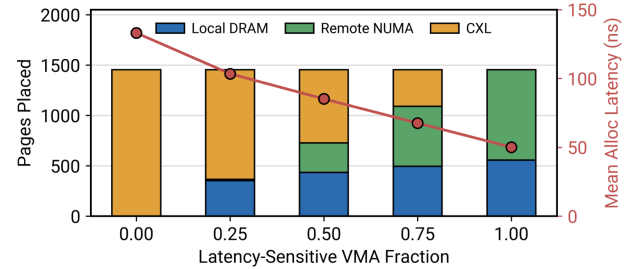}
  \caption{\revCQ{Tier-aware placement as latency-sensitive VMA fraction varies (simulation).
           Bars show pages placed per tier (left axis). The red line shows mean allocation latency (right axis).
           The per-VMA intent hint steers placement from CXL (fraction 0) to Local DRAM (fraction 1.0), cutting mean latency by $2.6\times$.}}
  \label{fig:tier-dist}
\end{figure}

\commonqnote{CQ3}\revCQ{\textbf{Tier-Aware Placement: Intent-Steered Allocation.}
Fig.~\ref{fig:tier-dist} shows the placement decisions and resulting mean allocation latency of the tier-aware library (\S\ref{sec:tier-aware}) as we vary the fraction of VMAs marked latency-sensitive (the per-VMA hint), with all other parameters fixed.
At a fraction of $0$, the utility objective pushes essentially all pages onto CXL, yielding ~130\,ns mean allocation latency.
As the fraction rises, placement shifts monotonically: CXL is displaced first by Remote NUMA, then by Local DRAM.
At fraction $1.0$, mean latency drops to ~50\,ns, a $2.6\times$ reduction.
Because the total pages placed is constant across all points, this improvement comes entirely from \emph{where} pages land.
Two points follow.
First, the per-VMA intent hint demonstrably steers placement. The library actively pulls hot data toward fast memory and pushes cold data off it, rather than greedily filling the fastest tier.
Second, this is inexpressible by a tier-blind fixed-function allocator, which has no mechanism to act on per-VMA software intent. Together, Figs.~\ref{fig:taa_results} and~\ref{fig:tier-dist} show the distinction that Fig.~\ref{fig:design_space} intentionally factors out. Fixed HW is excellent for one frozen policy. Valinor approaches that bound while also running adaptive policies on the same programmable substrate.}

\subsection{Area \& Power Overheads}
To quantify hardware cost, we used the Yosys Open SYnthesis Suite~\cite{yosys} to measure area and estimate power for both \textbf{Fixed-HW} and \textbf{Valinor}.
For \textbf{Valinor}, we synthesized the 3-stage in-order PAE, assuming an 8-core system (Small BOOM cores) with a single shared engine, using the nangate45 library \cite{nangate45}.
Fixed-HW adds only 0.12\% chip area, while Valinor incurs a modest 1.5\%. Static power scales with area, yielding a 1.5–1.8\% leakage overhead for \textbf{Valinor}, compared to <0.2\% for \textbf{Fixed-HW}.
Overall, Valinor's flexibility incurs minimal hardware overhead, thanks to a single unit shared system-wide.

\section{Related Work}
\label{sec::related_work}
To our knowledge, Valinor is the first scheme to combine hardware-class allocation performance with the flexibility and programmability of software-defined policies. 
In this section, we qualitatively compare Valinor to prior efforts that extend hardware or software to accelerate memory management.

\textbf{Hardware-Based Allocation.} Prior works \cite{tirumalasetty2022reducing, hbdpISCA2020, guo2022clio, pemberton2019enabling} accelerate page-fault handling by moving parts of the OS \vnit{handler} off the critical path, maintaining a hardware-managed pool of free pages~\cite{hbdpISCA2020}, or establishing mappings in hardware~\cite{guo2022clio}. While these approaches reduce fault latency, they (i) consume system resources for background tasks, (ii) offer no control over placement policy, and (iii) still rely on software to manage physical-page reservation.
In contrast, Valinor offloads the entire page-level allocation path to hardware, while allowing software-defined mapping policies.

\textbf{Offloading Memory Management to Hardware.} A large body of work accelerates memory management at \emph{object} granularity \cite{mementoMICRO2023, kanev2017, cam1999memalloc, vm23, vm24, vm21, vm22, wright2005objectaware, chang1993objectcaching}, often incorporating hardware-assisted or hardware-managed allocators and garbage collectors \cite{joao2009flexible, thomas2002gc, schmidt1994rtgc, wise1997rc, meyer2005onchipgc, garcia2021integratedgc, jang2019charon}. These systems require ISA extensions to notify hardware of allocation and free events, and focus on small, malloc-sized objects. Valinor differs fundamentally: it targets \emph{page-granularity} allocation and relies exclusively on existing MMU events (TLB misses and faults), avoiding ISA changes. As such, it is orthogonal to prior work on hardware-assisted object allocators and garbage collectors, which operate at much finer granularity and require explicit ISA support.

\section{Conclusion}
We presented Valinor, a hardware–OS cooperative allocation substrate that pairs software programmability with hardware-class latency. Valinor introduces a programmable hardware allocation engine (PAE) that executes OS-supplied allocation libraries, enabling low-latency page allocation while supporting diverse policies, from short-lived object allocation and integrity enforcement to telemetry-guided placement.
Implemented on a BOOM RISC-V soft core and evaluated in hardware and simulation, Valinor reduces allocation latency by 17×, improves end-to-end performance by 16\%, and cuts energy by up to 8\%, while supporting several allocation libraries.
Valinor demonstrates that memory allocation can be both efficient and fully programmable, enabling a new class of adaptable, policy-aware virtual memory systems.

\section*{Acknowledgments}
We thank the SAFARI Research Group members for their constructive feedback and for providing a stimulating intellectual and scholarly environment. We acknowledge the generous gift funding provided by our industrial partners (especially Google, Huawei, Intel, Microsoft), which has been instrumental in enabling the research we have been conducting on memory systems.


\bibliographystyle{unsrt}
\bibliography{refs}

\end{document}